\documentclass[journal]{IEEEtran}

\pdfoutput=1
\IEEEoverridecommandlockouts

\usepackage{amsmath}
\usepackage[english]{babel}
\usepackage{cite}
\usepackage[shortlabels,inline]{enumitem}
\usepackage{graphicx}
\usepackage{hyperref}
\usepackage{bookmark}
\usepackage{ifthen}
\usepackage[utf8]{inputenc}
\usepackage[locale=US,group-separator={,}]{siunitx}
\usepackage[T1]{fontenc}
\usepackage[subnum]{cases}
\usepackage[toc,symbols]{glossaries}
\usepackage{bm}
\usepackage{array}
\usepackage{tabularray}
\usepackage{acro}
\usepackage{microtype}
\usepackage{pifont}

\usepackage[utopia, cal=cmcal]{mathdesign} 
\DeclareSIUnit{\bitsperminute}{\text{bit} \, \text{min}^{-1}}
\newcommand{\xmark}{\ding{55}}
\newcommand{\cmark}{\ding{51}}%
\usepackage{subcaption}

\allowdisplaybreaks

\newcommand{\Figure}[1]{Fig.~\ref{#1}}
\newcommand{\Figures}[2]{Figs.~\ref{#1} and~\ref{#2}}
\newcommand{\Equation}[1]{\eqref{#1}}
\newcommand{\Table}[1]{Table~\ref{#1}}
\newcommand{\Section}[1]{Section~\ref{#1}}
\newcommand{\Sections}[2]{Sections~\ref{#1}~and~\ref{#2}}

\renewcommand{\vec}[1]{\mathbf{#1}}

\newcommand{\con}[0]{C_{\mathrm{GFPD}}} 
\newcommand{\volTube}[0]{V_{\mathrm{T}}} 

\newcommand{\Tg}[0]{T_{\mathrm{G}}} 
\newcommand{\veff}[0]{v_{\mathrm{eff}}} 
\newcommand{\D}[0]{D} 
\newcommand{\bit}[1]{b[#1]} 
\newcommand{\disTX}[0]{d_{\mathrm{TX,RX}}}

\newcommand{\nSymb}[0]{N_{\mathrm{Sym}}} 
\newcommand{\nBits}[0]{N_{\mathrm{Bit}}} 

\newcommand{\s}[0]{\text{s}} 

\DeclareMathOperator*{\argmax}{\mathrm{arg\,max}}

\newcommand{\n}[0]{n} 
\newcommand{\tn}[0]{t_n} 
\newcommand{\symIdx}[0]{k} 

\newcommand{\trainIdx}[0]{\tau}

\newcommand{\detecThreshIdx}[0]{j}
\newcommand{\detecThreshSetIdx}[0]{l}

\newcommand{\symVar}[0]{i}
\newcommand{\Ts}[0]{T_{\mathrm{S}}} 
\newcommand{\Ti}[0]{T_{\mathrm{I}}} 
\newcommand{\modOrder}[0]{M} 

\newcommand{\SR}[0]{s_{\mathrm{r}}}
\newcommand{\avgSR}[0]{\bar{s}_{\mathrm{r}}}
\newcommand{\recSig}[0]{r}
\newcommand{\sampleInt}[0]{\Delta t}

\newcommand{\ratioInten}[0]{\rho}
\newcommand{\txInten}[0]{I^{\mathrm{TX}}} 
\newcommand{\maxTxInten}[0]{I^{\mathrm{TX}}_{\mathrm{max}}} 

\newcommand{\trainLength}[0]{N_\mathrm{T}}
\newcommand{\trainSet}[0]{\mathcal{K}}
\newcommand{\trainMean}[0] {\hat{\mu}}
\newcommand{\trainVariance}[0]{\hat{\sigma}^2}
\newcommand{\trainThresh}[0]{\xi_\mathrm{T}}
\newcommand{\pFA}[0]{p_\mathrm{FA}} 
\newcommand{\tts}[0]{t_\mathrm{TS}} 

\newcommand{\tsEst}[0]{\hat{t}_\mathrm{s}}
\newcommand{\tsInit}[0]{\Tilde{t}_\mathrm{s}}

\newcommand{\searchInt}[0]{\mathcal{T}^{\mathrm{search}}}
\newcommand{\searchRadius}[0]{r}

\newcommand{\syncMetric}[0]{\Lambda}
\newcommand{\filterLength}[0]{\mathrm{N}}

\newcommand{\corrMarker}[0]{\mathrm{C}}
\newcommand{\diffMarker}[0]{\mathrm{D}}
\newcommand{\blindMarker}[0]{\mathrm{B}}

\newcommand{\filter}[0]{g}

\newcommand{\diffRecSig}[0]{\recSig_{\diffMarker}}

\newcommand{\blindCorrFilter}[0]{g_\corrMarker^{\blindMarker}}
\newcommand{\blindDiffCorrFilter}[0]{\filter_\diffMarker^\blindMarker}

\newcommand{\detecSample}[0]{d}
\newcommand{\estSym}[0]{\hat{{\symVar}}}
\newcommand{\detecThresh}[0]{\xi}
\newcommand{\detecThreshSet}[0]{\Xi}
\newcommand{\nPilots}[0]{P} 
\newcommand{\nCoherence}[0]{F} 
\newcommand{\nWindow}[0]{W} 
\newcommand{\nSkip}[0]{\chi} 

\newcommand{\symbolSampleSet}[0]{\mathcal{S}}
\newcommand{\avgSymbolSampleSet}[0]{\bar{S}}
\newcommand{\threshDiff}[0]{\delta}
\newcommand{\avgThreshDiff}[0]{\Delta}

\newcommand{\AMED}[0]{\mathrm{AMED}}

\usepackage{acro}

\DeclareAcronym{1D}{
    short = 1-D,
    long = one-dimensional
}
\DeclareAcronym{2D}{
    short = 2-D,
    long = two-dimensional
}
\DeclareAcronym{3D}{
    short = 3-D,
    long = three-dimensional
}
\DeclareAcronym{MC}{
    short = MC,
    long = molecular communication
}
\DeclareAcronym{OOK}{
    short = OOK,
    long = ON-OFF keying
}
\DeclareAcronym{ISI}{
    short = ISI,
    long = inter-symbol interference
}
\DeclareAcronym{IUI}{
    short = IUI,
    long = inter-user interference
}
\DeclareAcronym{IR}{
    short = IR,
    long = impulse response
}
\DeclareAcronym{ML}{
    short = ML,
    long = maximum likelihood
}
\DeclareAcronym{wlog}{
    short = w.l.o.g.,
    long = without loss of generality
}
\DeclareAcronym{BER}{
    short = BER,
    long = bit error rate
}
\DeclareAcronym{SER}{
    short = SER,
    long = symbol error rate
}
\DeclareAcronym{BSC}{
    short = BSC,
    long = Binary Symmetric Channel
}
\DeclareAcronym{CDF}{
    short = CDF,
    long = cumulative density function
}
\DeclareAcronym{UCA}{
    short = UCA,
    long = uniform concentration assumption
}
\DeclareAcronym{AWGN}{
    short = AWGN,
    long = Additive White Gaussian Noise
}
\DeclareAcronym{PBS}{
    short = PBS,
    long = particle-based simulation
}
\DeclareAcronym{MLSE}{
    short = MLSE,
    long = maximum likelihood sequence estimator
}
\DeclareAcronym{VE}{
    short = VE,
    long = viterbi equalizer
}
\DeclareAcronym{MLE}{
    short = MLE,
    long = maximum likelihood estimator
}
\DeclareAcronym{CIR}{
    short = CIR,
    long = channel impulse response
}
\DeclareAcronym{CIRs}{
    short = CIRs,
    long = channel impulse responses
}
\DeclareAcronym{SNR}{
    short = SNR,
    long = signal-to-noise ratio
}
\DeclareAcronym{TX}{
    short = TX,
    long = transmitter
}
\DeclareAcronym{TXs}{
    short = TXs,
    long = transmitters
}
\DeclareAcronym{RX}{
    short = RX,
    long = receiver
}
\DeclareAcronym{RXs}{
    short = RXs,
    long = receivers
}
\DeclareAcronym{ARE}{
    short = ARE,
    long = area rate efficiency
}
\DeclareAcronym{wrt}{
    short = w.r.t.,
    long = with respect to
}
\DeclareAcronym{GFP}{
    short = GFP,
    long = green fluorescent protein
}
\DeclareAcronym{GFPD}{
    short = GFPD,
    long = green fluorescent protein variant "Dreiklang"
}
\DeclareAcronym{EX}{
    short = EX,
    long = eraser
}
\DeclareAcronym{SM}{
    short = SM,
    long = signaling molecule
}

\DeclareAcronym{LB}{
    short = LB,
    long = lysogeny broth
}
\DeclareAcronym{OD600}{
    short = OD600,
    long = optical density at 600 nm
}
\DeclareAcronym{lacoperon}{
    short = lac operon,
    long = lactose operon
}
\DeclareAcronym{IPTG}{
    short = IPTG,
    long = Isopropyl-$\beta$-D-thiogalactopyranosid
}
\DeclareAcronym{IMAC}{
    short = IMAC,
    long = immobilised metal chelate affinity chromatography
}
\DeclareAcronym{CV}{
    short = CV,
    long = column volume
}
\DeclareAcronym{BCA}{
    short = BCA,
    long = bicinchoninic assay
}
\DeclareAcronym{SDSPAGE}{
    short = SDS-PAGE,
    long = sodium dodecyl sulfate polyacrylamide gel electrophoresis
}
\DeclareAcronym{YFP}{
    short = YFP,
    long = yellow fluorescent protein
}
\DeclareAcronym{FPs}{
    short = FPs,
    long = Fluorescent Proteins
}
\DeclareAcronym{RSFPs}{
    short = RSFPs,
    long = Reversible switchable fluorescent proteins
}
\DeclareAcronym{LED}{
    short = LED,
    long = light emitting diode
}
\DeclareAcronym{PWM}{
    short = PWM,
    long = pulse width modulation
}
\DeclareAcronym{FP}{
    short = FP,
    long = fluorescent protein
}
\DeclareAcronym{IoBNT}{
    short = IoBNT,
    long = Internet of Bio-Nano-Things
}
\DeclareAcronym{PBSbuffer}{
    short = PBS,
    long = phosphate buffered saline
}
\DeclareAcronym{ECOLI}{
    short = \textit{E.~coli},
    long = \textit{Escherichia coli}
}

\DeclareAcronym{DNA}{
    short = DNA,
    long = deoxyribonucleic acid
}
\DeclareAcronym{FEP}{
    short = FEP,
    long = fluorinated ethylene propylene
}
\DeclareAcronym{DOI}{
    short = DOI,
    long = digital object identifier
}

\DeclareAcronym{SPION}{
    short = SPION,
    long = superparamagnetic iron-oxide nanoparticle
}

\DeclareAcronym{UV}{
    short = UV,
    long = ultraviolet
}
\DeclareAcronym{AMED}{
    short = AMED,
    long = absolute mean Euclidean distance
}
\DeclareAcronym{NMSED}{
    short = NMSED,
    long = normalized minimum squared Euclidean distance
}

\DeclareAcronym{CS}{
    short = CS,
    long = correlation-based synchronization}

\DeclareAcronym{DCS}{
    short = DCS,
    long = differential correlation-based synchronization}

\DeclareAcronym{CSK}{
    short = CSK,
    long = concentration shift keying
}
\DeclareAcronym{LOWESS}{
    short = LOWESS,
    long = locally weighted scatterplot smoothing
}
\DeclareAcronym{SIR}{
    short = SIR,
    long = symbol impulse response
}
\DeclareAcronym{SR}{
    short = SR,
    long = signal response
}
\DeclareAcronym{SCF}{
    short = SCF,
    long = smoothed correlation filter
}
\DeclareAcronym{SDCF}{
    short = SDCF,
    long = smoothed differential correlation filter
}
\DeclareAcronym{BCF}{
    short = BCF,
    long = blind correlation filter
}
\DeclareAcronym{BDCF}{
    short = BDCF,
    long = blind differential correlation filter
}

\makeatletter
\long\def\@makecaption#1#2{\ifx\@captype\@IEEEtablestring%
    \footnotesize\begin{center}{\normalfont\footnotesize #1}\\
        {\normalfont\footnotesize\scshape #2}\end{center}%
    \@IEEEtablecaptionsepspace
    \else
    \@IEEEfigurecaptionsepspace
    \setbox\@tempboxa\hbox{\normalfont\footnotesize {#1.}~~ #2}%
    \ifdim \wd\@tempboxa >\hsize%
    \setbox\@tempboxa\hbox{\normalfont\footnotesize {#1.}~~ }%
    \parbox[t]{\hsize}{\normalfont\footnotesize \noindent\unhbox\@tempboxa#2}%
    \else
    \hbox to\hsize{\normalfont\footnotesize\hfil\box\@tempboxa\hfil}\fi\fi}
\makeatother

\begin{document}
\bstctlcite{IEEEexample:BSTcontrol}
\title{Closed-Loop Long-Term Experimental Molecular Communication System}
\author{
\IEEEauthorblockN{Maike Scherer$^\star$, Lukas Brand$^\star$, Louis Wolf, Teena tom Dieck, Maximilian Schäfer, Sebastian Lotter, Andreas Burkovski, Heinrich Sticht, Robert Schober, and Kathrin Castiglione\\
\thanks{$^\star$ Co-first authors.\\This paper was presented in part at the IEEE International Conference on Communications, 2024 \cite{brand2024closed}.}}
\IEEEauthorblockA{\small Friedrich-Alexander-Universit\"at Erlangen-N\"urnberg, Erlangen, Germany}}
\maketitle
\begin{abstract}
We present a fluid-based experimental molecular communication (MC) testbed which uses media modulation. Motivated by the natural human cardiovascular system, the testbed operates in a closed-loop tube system.
The proposed system is designed to be resource-efficient and controllable from outside the tube. As signaling molecule, the testbed employs the biocompatible green fluorescent protein variant "Dreiklang" (GFPD). GFPDs can be reversibly switched via light of different wavelengths between a bright fluorescent state and a less fluorescent state. GFPDs in solution are filled into the testbed prior to the start of information transmission and remain there for an entire experiment. For information transmission, an optical transmitter (TX) and an optical eraser (EX), which are located outside the tube, are used to write and erase the information encoded in the state of the GFPDs, respectively. At the receiver (RX), the state of the GFPDs is read out by fluorescence detection.
In our testbed, due to the closed-loop setup and the long experiment durations of up to 125 hours, we observe new forms of inter-symbol interferences (ISI), which do not occur in short experiments and open-loop systems. In particular, up to four different forms of ISI, namely channel ISI, inter-loop ISI, offset ISI, and permanent ISI, occur in the considered system.
For the testbed, we developed a communication scheme, which includes blind transmission start detection, symbol-by-symbol synchronization, and adaptive threshold detection, that supports higher order modulation.
We comprehensively analyze our MC experiments using the absolute mean Euclidean distance (AMED), eye diagram, and bit error rate (BER) as performance metrics.
Furthermore, we experimentally demonstrate the error-free transmission of $5,370$ bit at a data rate of 36 $\textrm{bit}\, \textrm{min}^{\boldsymbol{-1}}$ using 8-ary modulation and the error-free binary transmission of around $90,000$ bit at a data rate of 12 $\textrm{bit}\, \textrm{min}^{\boldsymbol{-1}}$. For the latter experiment, data was transmitted continuously for a period of more than five days (125 hours) during which no signaling molecules were injected into or removed from the system.
All signals recorded during the experiments, representing more than 250 kbit of data transmitted via MC, and parts of the evaluation code are publicly available on Zenodo and Github, respectively.
\end{abstract}
\acresetall
\section{Introduction}\label{sec:intro}
Synthetic \ac{MC} is an emerging research field at the intersection of biology, nanotechnology, and communications engineering. Inspired by natural processes, \ac{MC} offers innovative methods to transmit information by encoding it into chemical signals \cite{nakano2013molecular}. This approach has the potential to enable transformative applications in medicine \cite{felicetti2016applications,akyildiz2015internet}, nanotechnology \cite{Akan2012nanonetworks,soldner2020survey}, agriculture \cite{Dixon1990agricultural}, and environmental monitoring \cite{nakano2012molecular} by providing biocompatible and energy-efficient communication solutions, particularly in environments where conventional communication methods are impractical \cite{farsad2016comprehensive}.

So far, most work in \ac{MC} has focused on theoretical research, resulting in the development of theoretical models and communication schemes \cite{jamali2019channel,kuscu2019transmitter,kuran2020survey, farsad2016comprehensive}.
However, in order to advance from theory to the envisioned applications of \ac{MC}, it is crucial to develop experimental testbeds to bridge the gap between theoretical concepts and their practical application \cite{lotter2023experimental}. 
In recent years, there has been a notable increase in experimental \ac{MC} research and the number of experimental testbeds. A comprehensive overview on experimental \ac{MC} is provided in \cite{lotter2023experimental, Lotter2023testbedII}. 

A significant portion of \ac{MC} research has focused on medical applications, including in-body communication, health monitoring, and targeted drug delivery \cite{felicetti2016applications}. Consequently, suitable \ac{MC} testbeds aim to model these application scenarios, e.g., \ac{MC} within the human cardiovascular system, which directly influences their design. Therefore, many existing \ac{MC} testbeds \cite{bartunik2023development,lin2024testbed,wang2020understanding,wietfeld2024evaluation,angerbauer2023salinity,farsad2017novel,walter2023real} try to emulate real-world conditions, e.g., by employing biocompatible signaling molecules and by mimicking relevant aspects of the closed-loop human cardiovascular system using tube-based propagation channels with a background fluid flow \cite{Lotter2023testbedII}. However, most existing testbeds are only partially successful in this regard. On the one hand, the testbeds presented in \cite{bartunik2023development,lin2024testbed,wang2020understanding} use biocompatible signaling molecules -- specifically, \acp{SPION}, the cyanine dye indocyanine green, and the salt sodium chloride, respectively, that have the potential to be used in future \textit{in vivo} applications. On the other hand, all previously mentioned testbeds \cite{bartunik2023development,lin2024testbed,wang2020understanding,wietfeld2024evaluation,angerbauer2023salinity,farsad2017novel,walter2023real} share the limitation of considering only simple topologies -- primarily a single straight duct as the propagation channel -- and lack self-containment, i.e., they operate as open-loop systems. Here, open-loop refers to systems where the background fluid used in the experiment can enter or leave the system during operation. As a result, in these testbeds, the signaling molecules are added to the tube system at one point, e.g., via an injection, are used \textit{once} for information transmission, and then are collected as waste at the end of the tube. \textbf{Such systems have two major drawbacks}: First, long-term transmission experiments generate a lot of waste. Second, many of the intended applications target closed-loop environments, e.g., the human cardiovascular system, and cannot be accurately emulated by open-loop topologies.

Therefore, experimental \ac{MC} systems are required where the communication system operates within a closed-loop tube system. Currently, there are only two \ac{MC} testbeds operating in a closed-loop topology \cite{tuccitto2017fluorescent,schafer2024chorioallantoic}. In \cite{tuccitto2017fluorescent}, a fluid is pumped in a closed-loop system while fluorescent particles are injected, detected, and then diluted. The testbed presented in \cite{schafer2024chorioallantoic} constitutes the first \textit{in vivo} \ac{MC} testbed and is based on the chorioallantoic membrane of fertilized chicken eggs, where the fluorescent dye indocyanine green is injected into a closed-loop vascular system. However, the repeated dilution of fluorescent particles in \cite{tuccitto2017fluorescent} reduces their detectability in long-term experiments, and in \cite{schafer2024chorioallantoic}, only a single injection is considered due to the strong soiling of the vascular system by indocyanine green molecules. Moreover, in both testbeds, the signaling molecules can be used only for a single transmission, which is not resource-efficient. Therefore, although the testbeds in \cite{tuccitto2017fluorescent, schafer2024chorioallantoic} consider a closed loop, their applicability for long-term experiments is limited.

In this paper, we propose the first experimental closed-loop \ac{MC} testbed in which reversibly switchable signaling molecules are reused multiple times, enabled by molecular \textit{media modulation} \cite{Brand2022MediaModulation}. In media modulation, signaling molecules with specific switching properties \cite{brand2023switchable} are employed and, unlike in conventional \ac{MC} systems, they are injected only once into the system. Then, the state of the signaling molecules is repeatedly switched to transmit information, i.e., no additional molecules are injected into or removed from the system during operation. Compared to common injection-based release mechanisms \cite{lotter2023experimental}, in media modulation, the \ac{TX} can be placed outside the \ac{MC} channel and does not affect the propagation environment. As signaling molecule, we adopt the biocompatible \ac{GFPD} \cite{richards2003safety, brakemann2011reversibly}. \acp{GFPD} can be reversibly switched between a bright fluorescent ON state and a less fluorescent OFF state via light stimuli of different wavelengths \cite{brakemann2011reversibly}. This allows for writing and erasing information employing an optical \ac{TX} and \ac{EX}, respectively. The state of the \acp{GFPD}, which conveys the transmitted information, is read out via fluorescence detection at the \ac{RX}. In summary, the proposed \ac{MC} testbed facilitates resource-efficient communication and long-term experiments, and significantly reduces channel soling. By considering a closed-loop propagation environment, the testbed mimics an important property of the cardiovascular system, and the use of the biocompatible signaling molecule \ac{GFPD} ensures relevance for future applications. 

In this paper, we provide a comprehensive description and analysis of the proposed closed-loop media modulation testbed, its construction, and its components. Moreover, we discuss the chemical properties of the biocompatible signaling molecule \ac{GFPD} and their impact on communication performance. We identify three forms of \ac{ISI}, i.e., inter-loop \ac{ISI}, offset \ac{ISI}, and permanent \ac{ISI}, which occur particularly in closed-loop systems and during long-term experiments based on media modulation, and propose methods for their mitigation. In addition, we develop and analyze a suitable communication scheme for the considered experimental \ac{MC} system. Several communication schemes, including methods for modulation \cite{kuran2020survey,wietfeld2024evaluation}, synchronization \cite{lin2016time, jamali2017symbol, Xuewen2024synchro,Debus2024synchro}, as well as detection \cite{khaloopour2019experimental,kuscu2021fabrication,walter2023real}, have been developed for \ac{MC}. In this paper, we desgin a communication scheme suitable for the proposed closed-loop media modulation testbed, building on the existing literature. In particular, the developed scheme comprises a noise estimation-based wake-up method, higher-order modulation, blind and data-based synchronization, and adaptive threshold-based detection. Although these methods are developed with a focus on the considered media modulation testbed, they can be generalized to other experimental \ac{MC} testbeds, as they are deliberately introduced in a general form. Furthermore, we discuss relevant qualitative and quantitative performance metrics for experimental \ac{MC} systems, including \ac{BER}, eye diagram, and \ac{AMED}, and apply these metrics to evaluate the performance of the developed communication scheme.
Furthermore, we present the first long-term \ac{MC} experiment, achieving error-free transmission of around $90,000 \, \si{bit}$ sent over more than five days. This demonstrates the reliability of the proposed testbed and highlights the potential of media modulation-based \ac{MC} as a promising approach for future applications, such as long-term health monitoring. Finally, we compare the proposed media modulation testbed to existing \ac{MC} testbeds, in terms of communication-theoretical metrics such as the \ac{BER}, molecule efficiency as a relevant form of resource efficiency, and detection methods.

The main contributions of this paper can be summarized as follows:
\begin{itemize}
    \item We utilize light-based media modulation to transmit information in a closed-loop tube system using the states of the biocompatible and reusable signaling molecule \ac{GFPD}. This form of media modulation enables the operation of \ac{TX}, \ac{EX}, and \ac{RX} outside of the tube. As a result, they do not interfere with the propagation of the signaling molecules inside the tube; a feature that is beneficial, e.g., for healthcare applications. By considering a closed-loop topology, we ensure that the proposed testbed resembles and allows to study an important property of the cardiovascular system.
    \item Because of the closed-loop and long experiments, new forms of \ac{ISI} occur, which become apparent on different time scales. We characterize these forms of \ac{ISI} and implement suitable \ac{ISI} mitigation schemes. 
    \item We develop a communication scheme for the considered system that can also be used in other experimental \ac{MC} testbeds. The scheme includes a noise-estimation based wake-up method and data-based and blind synchronization. Furthermore, for detection, we propose an adaptive threshold detector. The transmission characteristics of the testbed are evaluated comprehensively - specifically by metrics such as \ac{AMED}, eye diagrams, and \ac{BER}.
    \item We experimentally demonstrate the error-free data transmission of $5,370 \, \si{bit}$ at a data rate of $36 \, \si{\bitsperminute}$, where 8-ary modulation is used. In addition, to the best of our knowledge, we report the longest MC experiment to date: Error-free transmission of around $90,000 \, \si{bit}$ over a period of more than $5$ consecutive days ($125 \, \si{\hour}$) at a data rate of $12 \, \si{\bitsperminute}$. No molecules were injected into or extracted from the system during this time, i.e., only $9 \, \si{\milli \liter}$ of \ac{GFPD} solution with a concentration of $\con = 0.3 \,\si{\milli\gram \per\milli\liter}$ were required for the entire experiment. By this, we  demonstrate the superiority of media modulation for long \ac{MC} experiments compared to the repeated injection of molecules employed in existing \ac{MC} testbeds.
    \item All our experimental data (more than 250 kbit transmitted across different experimental settings) as well as the Python code for evaluation have been released in open access \cite{scherer2025Zenodo}. This large amount of experimental data enables other researchers to develop and evaluate alternative processing methods for the received signals, e.g., machine learning based algorithms.
\end{itemize}
We note that a preliminary and simplified version of the testbed was briefly presented as part of \cite{brand2023switchable}. In \cite{brand2023switchable}, the experiment duration was only $3 \,\si{\minute}$, which did not allow to i) evaluate the resource efficiency of the testbed, ii) characterize the \ac{ISI}, and iii) experimentally determine the \ac{BER} for different detection schemes, which are new contributions of this work. Moreover, this paper significantly extends its conference version \cite{brand2024closed}: We extend the communication schemes developed in \cite{brand2024closed} by methods for synchronization, blind transmission start estimation, and adaptive threshold detection. Moreover, in this paper, we consider higher order modulation schemes, i.e., 8-ary modulation, and employ the \ac{AMED} and eye diagrams as performance metrics to evaluate the transmission characteristics of the testbed. While in \cite{brand2024closed}  the error free transmission of $500\, \si{bit}$ for an experiment of duration $250 \,\si{\minute}$ was shown, in this paper, we consider the transmission of $90,000\,\si{bit}$ for over $5$ days. 

The remainder of this paper is organized as follows. In \Section{sec:experiment}, we provide an overview of the proposed closed-loop testbed, including a detailed description of the signaling molecule \ac{GFPD}, the testbed components, and the working principle of the testbed. We discuss the physical properties of the testbed and their influence on communication performance in \Section{sec:characterization}. In \Section{sec:communication}, we describe the proposed communication scheme in detail, including modulation, symbol synchronization, and detection. The proposed performance metrics are presented in \Section{sec:metric}, and our results and evaluation are provided in \Section{sec:results}, which we compare with results achieved with other testbeds in \Section{sec:comparison_Testbeds}. Finally, \Section{sec:conclusion} concludes the paper and outlines topics for future work.
\section{Testbed Overview}\label{sec:experiment}
\begin{figure*}[!tbp]
    \begin{subfigure}[t]{0.99\textwidth} 
        \caption{}
        \centering
        \includegraphics[width=\textwidth]{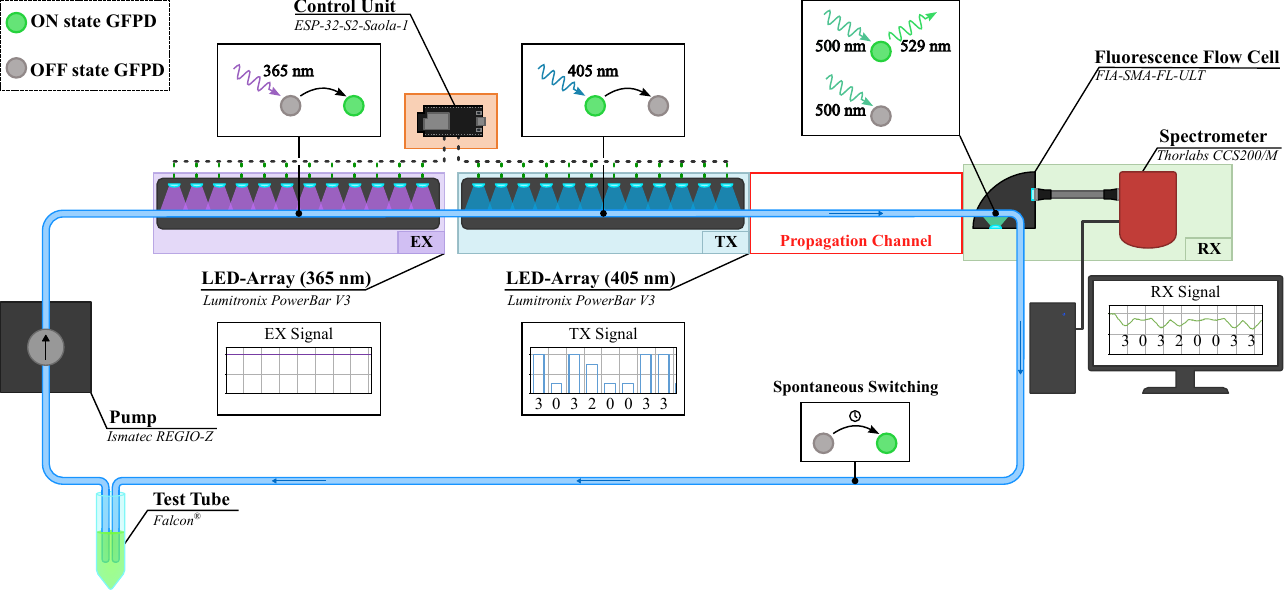}
        \label{fig:overview} 
    \end{subfigure}
    \begin{subfigure}[t]{0.8\textwidth} 
        \caption{}
        \includegraphics[height=6cm]{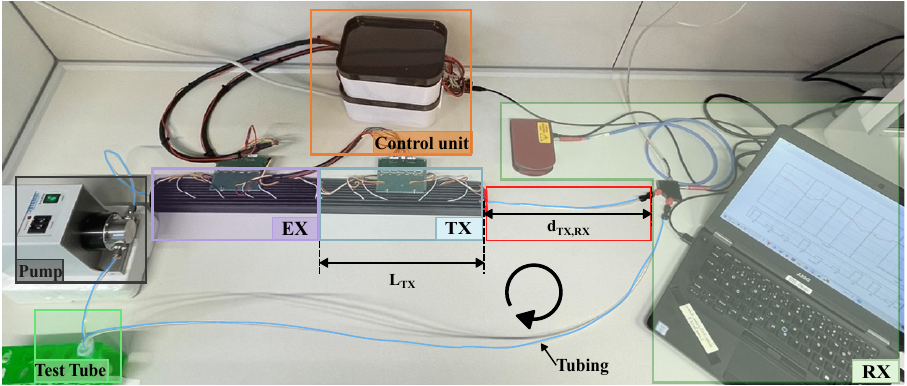}        
        \label{fig:lab_overview}
    \end{subfigure}
    \hspace{0.05\textwidth} 
    \begin{subfigure}[t]{0.1\textwidth}
        \caption{}
        \includegraphics[height=6cm]{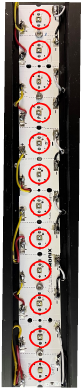}
        \label{fig:led_array}
    \end{subfigure}
    \caption{(a) Schematic representation of the testbed consisting of the propagation channel, pump, test tube, two LED-arrays serving as \ac{TX} and \ac{EX}, respectively, a flow cell, and a spectrometer acting as \ac{RX}. The switching processes of \ac{GFPD} are depicted next to the components in which they occur. (b) Photo of the experimental setup during operation. (c) \ac{LED} array of the \ac{TX} module (the \ac{LED} module of the \ac{EX} looks similar). Individual \acp{LED} of the array are highlighted by red circles.}
    \label{fig:total_overview}
\end{figure*}

In this section, an overview of the testbed is provided.
A schematic representation of the building blocks of the testbed and their functionality is displayed in \Figure{fig:overview}, while \Figure{fig:lab_overview} shows the testbed during operation. 
\subsection{Closed-Loop System: Tube System and Pump}\label{subsec:structure}
For our testbed, we use a \ac{FEP} tube with radius $r_\mathrm{T} = 8 \, \times 10^{-4} \, \si{\meter}$ and total length $L_\mathrm{T} = 2.74 \,\si{\meter}$. To achieve the closed-loop structure, both ends of the tube are put into a test tube (\textit{Falcon}$\,\textsuperscript{\tiny\textregistered}$). The \textit{Falcon}$\,\textsuperscript{\tiny\textregistered}$ is filled with $V_\mathrm{R} = 2.50 \,\si{\milli\liter}$ of fluid\footnote{In fact, the fluid is a solution of dissolved \ac{GFPD}. Details on the composition of the \ac{GFPD} solution and its introduction into the system are provided in \Section{SubSec:Procedure}.}. As long as both ends of the tube are immersed in the fluid of the \textit{Falcon}$\,\textsuperscript{\tiny\textregistered}$, they are hereby connected. To ensure that both tube ends are immersed in the fluid even if the tubes move slightly during the pumping process, we found $V_\mathrm{R} = 2.50 \,\si{\milli\liter}$ to be a suitable amount.

For pumping, a gear pump (\textit{Ismatec REGLO-Z}), attached to the tube, is used. The pump drives the fluid in the system with an effective volume flux of $Q = 9.45 \, \mathrm{mL}\;\mathrm{min}^{-1}$, leading to an effective flow velocity of $\veff = 0.139 \,\si{\meter \per\second}$ for the given radius of the tube $r_\mathrm{T}$. The closed-loop pipe system holds a total volume of $V_{\mathrm{tot}} = 9 \,\si{\milli\liter}$ of liquid, cf. \Section{SubSec:Procedure} for details. Thus, the average time that the fluid needs to circulate one time through the testbed loop is $T_\mathrm{L} = \frac{V_{\mathrm{tot}}}{Q} = 57 \,\si{\second}$. This also implies that, on average, every $\frac{V_\mathrm{R}}{Q} = 15\,\si{\second}$ the fluid in the \textit{Falcon}$\,\textsuperscript{\tiny\textregistered}$ is fully replaced.
\subsection{Signaling Molecule: GFPD and Its Properties}\label{subsec:GFPD}
\ac{GFPD} is a monomeric photoswitchable and biocompatible \ac{GFP} introduced in \cite{brakemann2011reversibly}. The authors in \cite{brakemann2011reversibly} show that \ac{GFPD} has some unique properties, which make it well suited for the use in our testbed. In particular, the switching efficiency, robustness to photobleaching, and fluorescence intensity are high for \ac{GFPD}. In the following, we discuss these properties of \ac{GFPD} in detail. 
\subsubsection{Reversible Photoswitchability -- Modulation und Resetting via Active Switching}\label{par:switching}
In general, \textit{photoswitchability} means that the properties of a molecule can be changed upon exposure to light of specific wavelengths. In the case of \ac{GFPD}, photoswitchability allows for a reversible switching of its fluorescence between two\footnote{In fact, a third state exists, which we denote as \textit{equilibrium state}. In the absence of light, \ac{GFPD} may switch to this state, which shows a fluorescence similar to that of the ON state. As the equilibrium state's fluorescence is only slightly higher, in this work, for simplicity of presentation, we do not differentiate between the ON state and the equilibrium state.} distinct states, namely the ON and OFF states, which can be controlled by irradiation of light with different wavelengths.
In particular, \ac{GFPD} may be switched to the fluorescence ON and OFF states upon irradiation with light of wavelengths $\lambda_\mathrm{ON} = 365 \, \si{\nm}$ and $\lambda_\mathrm{OFF} = 405 \, \si{\nm}$, respectively. On the molecular level, the irradiation triggers the dehydration ($\lambda_\mathrm{ON} = 365 \, \si{\nm}$) and hydration ($\lambda_\mathrm{OFF} = 405 \, \si{\nm}$) of the imidazolinone ring, i.e., one of two rings the chromophore consists of, which leads to a shift in the absorbance spectrum, cf. \cite[Fig. 1e)]{brakemann2011reversibly}. The success of the switching depends on the intensity of the irradiation.
The switching property allows the use of \ac{GFPD} molecules as rewritable\footnote{In practice, data storage units often have limited program-erase cycles, i.e., they can only be overwritten a finite number of times before they wear out, as is known, e.g., for flash memory devices \cite{bez2003introduction}. Similarly, \ac{GFPD}, due to its biological nature, also has a finite number of program-erase cycles until photobleaching occurs, which is irradiation dependent, cf. \Section{par:photobleaching}. This phenomenon has also been observed for other switchable fluorescent proteins~\cite{duan2013structural}. An endurance analysis of \ac{GFPD} was presented in~\cite[Suppl. Fig. 4]{brakemann2011reversibly}, showing that the fluorescence intensity is reduced by half after approximately 190 switching cycles, i.e., program-erase cycles. However, the irradiation intensities used for switching in~\cite[Suppl. Fig. 4]{brakemann2011reversibly} are higher than those applied in our work. Therefore, we expect that a higher number of program-erase cycles is possible in our testbed compared to \cite[Suppl. Fig. 4]{brakemann2011reversibly}.} information storage units. In particular, the switching property enables the modulation and deletion of information into and from the state of the \ac{GFPD}, facilitated by a \ac{TX} and \ac{EX} module in our testbed, respectively, which are discussed in detail in \Section{SubSec:InfComp}.
\subsubsection{Fluorescence-Based State Readout}\label{par:readout}
Only \acp{GFPD} in the ON state show a high fluorescence, i.e., emit light at a wavelength of $\lambda_\mathrm{E} = 529 \, \si{\nm}$ after being excited by light of wavelength $\lambda_\mathrm{T} = 500 \, \si{\nm}$. Hence, the wavelengths responsible for switching, i.e., $365 \, \si{\nm}$ and $405 \, \si{\nm}$, are decoupled from the fluorescence excitation and detection wavelengths. To read the fluorescence, we use a spectrometer in our testbed, which is detailed in \Section{SubSec:InfComp}.
\subsubsection{Thermal Relaxation to the ON State}\label{par:relaxation}
Thermal relaxation denotes the process by which a perturbed system spontaneously returns to its equilibrium state. For \ac{GFPD}, this means that an OFF state \ac{GFPD} can spontaneously switch back to its ON state. The thermal relaxation of \ac{GFPD} to the ON state has a half-life of $T_{1/2} = 600 \,\si{\second}$ at room temperature\footnote{Conducting experiments at $37\,\si{\degreeCelsius}$, which corresponds to the human body temperature, may be more realistic for biological in-body applications. However, such experiments are significantly more complex, as they require specialized equipment such as a heating chamber. Therefore, we conducted only a few individual experiments at $37\,\si{\degreeCelsius}$ to verify that communication remains feasible at human body temperature. The results confirmed that reliable communication is maintained at $37\,\si{\degreeCelsius}$. However, as expected, the received signal appeared noisier at $37\,\si{\degreeCelsius}$ compared to $21\,\si{\degreeCelsius}$. This observation is consistent with \cite{brakemann2011reversibly}, where a decrease in the maximum fluorescence intensity with increasing temperature was reported for \ac{GFPD}, thereby increasing the relative impact of measurement noise. Additionally, the experiments at $37\,\si{\degreeCelsius}$ showed an accelerated thermal relaxation process, which is expected, as a half-life of approximately $T_{1/2} = 150\,\si{\second}$ at $37\,\si{\degreeCelsius}$ has been reported in \cite{brakemann2011reversibly}. Due to space constraints, only the results for room temperature experiments are included in this paper.}~\cite{brakemann2011reversibly}. As the half-life $T_{1/2}$ depends on temperature, cf. \cite[Fig. 1f)]{brakemann2011reversibly}, a laboratory with constant room temperature, $21 \, \si{\celsius}$ in our case, supports reproducible experimental results.

We note that only the ON state of \ac{GFPD} is thermodynamically stable \cite{brakemann2011reversibly}, i.e., the switching of \ac{GFPD} from the ON state to the OFF state without a light trigger is not expected. Consequently, the ON state is predominant in equilibrium in the proposed testbed.
\subsubsection{Photobleaching}\label{par:photobleaching}
After multiple photoswitching cycles and readouts, photobleaching of \ac{GFPD} may occur \cite{brakemann2011reversibly}. Photobleaching is a general phenomenon when irradiating \ac{GFP} and its variants such as \ac{GFPD} and refers to the rare event of destruction of the fluorophore molecule upon exposure to light. Photobleaching is energy dependent, i.e., it increases with higher irradiation power or longer irradiation duration \cite[Suppl. Fig. 7]{brakemann2011reversibly}.
A photobleached \ac{GFPD} irreversibly loses its ability to be fluorescent \cite{wang2016gmars, brakemann2011reversibly}, which can be interpreted as \ac{GFPD} degradation, visible as a gradual decrease in fluorescence intensity of \ac{GFPD} solutions upon repeated exposure to light.
Additionally, while the aggregation of intact \acp{GFPD} at low protein concentrations in an adequately buffered system has not been reported as a significant issue, the aggregation of photobleached \acp{GFPD} is expected. In particular, photobleaching is often a result of the oxidation of proteins, which renders their tertiary structure thermodynamically unstable. As a consequence, oxidized proteins such as photobleached \ac{GFPD} tend to expose hydrophobic amino acids to the outside, which leads to agglomeration and cross-linking between proteins \cite{bosshard2010protein}. Moreover, the exposed hydrophobic regions result in an increased affinity for hydrophobic surfaces~\cite{suelter1983prevent}. In our case, this may cause aggregation of degraded \ac{GFPD} at the inner tube wall and at other system components.
\subsubsection{Biocompatibility}\label{sec:GFPD_bio}
Like all \acp{GFP}, \ac{GFPD} is \textit{biocompatible} \cite{richards2003safety}. This property allows the use of \ac{GFP}-based proteins in a wide range of living organisms without causing significant harm or disruption to their normal functions \cite{richards2003safety}. The biocompatibility of \ac{GFPD} also makes our laboratory experiments safe.
\subsection{Experimental Procedure} \label{SubSec:Procedure}
Before the testbed can be used for information transmission, the \ac{GFPD} solution must be prepared and filled into the testbed. Hence, each experiment involves the following steps.
\subsubsection{Step 1 (Preparation of GFPD Solution)} \label{SubSec:GFPD_Solution}
Phosphate-buffered saline (PBS) buffer, a water-based salt solution, is used as the liquid buffer medium. Specifically, a high-concentration stock of \ac{GFPD} is diluted with PBS buffer and the detergent Triton X-100, which is also needed to stabilize \ac{GFPD} for long-term experiments, to a total volume of $9 \,\si{\milli\liter}$. In the end, the \ac{GFPD} solution contains a \ac{GFPD} concentration of $\con = 0.3 \,\si{\milli\gram \per\milli\liter}$ and a Triton X-100 concentration of $C_{\mathrm{TRITON}} = 0.125 \,\si{\milli\gram \per\milli\liter}$.  Using the molecular weight of \ac{GFPD} of $ M_{\mathrm{w}} = 26.9 \,\si{\kilo\dalton} = 26900 \,\si{\gram\per\mol}$\cite{databaseDreiklang} and the Avogadro constant $ \mathrm{N}_{\mathrm{A}} = 6.02214076 \times 10^{23} \, \si{\per \mol}$, the number of \ac{GFPD} molecules in the system, $N_{\mathrm{GFPD}}$, can be estimated as $N_{\mathrm{GFPD}} = \frac{\con V_{\mathrm{tot}} \mathrm{N}_{\mathrm{A}}}{M_{\mathrm{w}}} = 6.04 \times 10^{16}$.
\subsubsection{Step 2 (Filling the GFPD Solution into the Closed-Loop System)} \label{SubSec:Filling}
The \ac{GFPD} solution is pumped into the tube, which has a volume of $\volTube = 5.43 \,\si{\milli\liter}$, and all segments within the pump and flow cell, which have a joint volume of $V_\mathrm{PF} = 1.07 \,\si{\milli\liter}$, until the test tube is filled with $V_\mathrm{R} = 2.50 \,\si{\milli\liter}$, which requires in total $9 \,\si{\milli\liter}$ of \ac{GFPD} solution. At this point, both ends of the tubing are placed in the test tube to close the loop.
\subsubsection{Step 3 (Information Transmission)} \label{SubSec:Transmission}
While the pump constantly transports the \ac{GFPD} solution in the loop, information can be modulated from outside the tube into the state of the \acp{GFPD} by an \ac{LED}-based \ac{TX}. Later, the \ac{RX} reads out the state of the \acp{GFPD} via fluorescence triggered by an \ac{LED}. Additionally, an \ac{LED}-based \ac{EX} can be used to erase the modulated information. The components that facilitate information transmission, i.e., the \ac{TX}, the \ac{TX}-\ac{RX} link, which we refer to as channel, the \ac{RX}, and the \ac{EX} are described next.
\subsection{Information Transmission Components} \label{SubSec:InfComp}
\subsubsection{The Transmitter - Optical-to-Chemical-Signal Conversion}\label{subsubsec:tx}
The \ac{TX} is used to modulate information via media modulation into the state of the \acp{GFPD}. Since the ON state is the idle state of the \acp{GFPD}, the \ac{TX} switches \ac{GFPD} to the OFF state to convey information.
The \ac{TX} consists of an \ac{LED} array with $12$ equally spaced \acp{LED} (\textit{PowerBar V3, Lumitronix}), cf. \Figure{fig:led_array}, which can emit light at a wavelength of $\lambda_\mathrm{OFF} = 405 \, \si{\nano\meter}$. The array is attached to an aluminum circuit board with a total length of $L_\mathrm{TX} = 29 \,\si{\cm}$ that is connected to a passive cooling element to increase the service life of the \acp{LED}. In addition, the \acp{LED} are individually wired for maximum flexibility. This design enables control of the irradiation intensity of each \ac{LED} by a microcontroller (\textit{ESP32-S2, Espressif Systems}) during the experiments. In particular, this allows us to vary the length of the \ac{TX} between $L_\mathrm{TX} = 2.5\,\si{\cm}$ (using only one \ac{LED}) up to $L_\mathrm{TX} = 29\,\si{\cm}$ (using all $12$ \acp{LED}). Since the switching probability of \ac{GFPD} depends on the energy supplied, increasing the number of \acp{LED} generally improves the information modulation process and therefore the transmission reliability\footnote{Note that, to ensure independent symbols can be generated at the \ac{TX}, its length cannot be increased arbitrarily. In particular, to ensure independent transmit symbols, the majority of excited \ac{GFPD} molecules must exit the \ac{TX} region before the next symbol interval begins. Otherwise, \ac{ISI} occurs already at the \ac{TX}.}. This energy dependency has been studied in detail in~\cite{Brand2022MediaModulation}, where the photoswitching behavior of \ac{GFPD} was modeled using the Beer–Lambert law~\cite[Eq. 6]{Brand2022MediaModulation}, and the influence of varying light power levels was analyzed comprehensively~\cite[Fig. 5]{Brand2022MediaModulation}. In this work, for simplicity, we focus on the case where all 12 \acp{LED} are switched on and off simultaneously, i.e., $L_\mathrm{TX} = 29\,\si{\cm}$. The \ac{TX} is placed on top of the tubing, cf. \Figure{fig:total_overview}.

To encode information, the \ac{TX} emits light for a duration of $\Ti$ and hereby controls the ON state concentration of the irradiated \ac{GFPD} molecules in the tubing, i.e., it controls the fraction of \ac{GFPD} molecules switched from the ON state to the OFF state. The irradiation intensity is adapted according to the symbol to be transmitted (cf. \Section{Sec:Modulation} for mathematical details). The switching of \ac{GFPD} is an energy-dependent probabilistic process \cite{brakemann2011reversibly, Brand2022MediaModulation}. In our testbed, the number of \acp{LED} employed, the irradiation duration $\Ti$, and the irradiation intensity jointly determine the switching probability. This is because they determine the total number of photons released during $\Ti$, which in turn determines the probability that a photon will hit a \ac{GFPD} molecule and thus trigger a switch.
$\Ti$ is followed by a guard interval of duration $\Tg$, during which the \ac{TX} is always inactive. This results in a symbol duration of $\Ts = \Ti + \Tg$.
\subsubsection{The Channel - Pump-Driven Advection}\label{subsubsec:channel}
In the testbed, information is transmitted over the communication channel of length $d_{\mathrm{TX},\mathrm{RX}}$, which we define to be the tubing between the end of the \ac{TX} and the beginning of the \ac{RX} (framed in red in \Figures{fig:overview}{fig:lab_overview}). In this work, two channel lengths are investigated, $d_{\mathrm{TX},\mathrm{RX}} = 6 \, \si{\centi\meter}$ and $d_{\mathrm{TX},\mathrm{RX}} = 35 \, \si{\centi\meter}$. Note that our testbed design allows for easy modification of $d_{\mathrm{TX},\mathrm{RX}}$ by moving the \ac{TX}, which is loosely placed on top of the tube. In principle, $d_{\mathrm{TX},\mathrm{RX}}$ could even be varied dynamically during experiments. In this work, however, the \ac{TX} is not moved during experiments.

During the propagation from the \ac{TX} to the \ac{RX}, modulated \ac{GFPD} molecules may spontaneously switch back to their ON state due to thermal relaxation.
In our experiments, the effect on the received signal is small as $T_{1/2} = 600 \,\si{\second}$ is large compared to the average time required to propagate the distance $d_{\mathrm{TX},\mathrm{RX}}$ ($0.4 \, \si{\second}$ and $2.5 \,\si{\second}$ for $d_{\mathrm{TX}, \mathrm{RX}} = 6 \, \si{\centi\meter}$ and $d_{\mathrm{TX},\mathrm{RX}} = 35 \, \si{\centi\meter}$, respectively, assuming $\veff = 0.139 \,\si{\meter \per\second}$). However, for future experiments with a longer channel, thermal relaxation may become more relevant.
\subsubsection{The Receiver - Fluorescence-Based Readout}\label{subsubsec:rx}
The \ac{RX} is used to read out the current states of the \acp{GFPD}. 
As \ac{RX}, a fluorescence flow cell (\textit{FIA-SMA-FL-ULT, Ocean Optics}), equipped with an \ac{LED} that can emit light at a wavelength of $\lambda_\mathrm{T} = 500 \, \si{\nm}$, is placed at a distance of $d_{\mathrm{TX},\mathrm{RX}}$ from the end of the \ac{TX}. The \ac{LED} is used to trigger the fluorescence of the \ac{GFPD}. The light emitted by the \acp{GFPD} upon fluorescence, which has a wavelength of $\lambda_\mathrm{E} = 529 \, \si{\nm}$, is guided by an optical fiber and measured by a compact spectrometer (\textit{CCS 100M, Thorlabs}). The measured signal is recorded by the \textit{ThorSpectra software}. 
\subsubsection{The Eraser - Deleting the Modulated Information}\label{subsubsec:ex}
The \ac{EX} can be used to reset the state of the \acp{GFPD} to the ON state. 
The \ac{EX} consists of an \ac{LED} array with $12$ equally spaced \acp{LED} (\textit{PowerBar V3, Lumitronix}), which can emit light at a wavelength of $\lambda_\mathrm{ON} = 365 \, \si{\nano\meter}$. Apart from the type of \acp{LED} used, the \ac{EX} is constructed and wired identical to the \ac{TX}, i.e., the hardware design and length of the \ac{EX} are identical to those of the \ac{TX}. Hence, the length of the \ac{EX} can be varied between $L_\mathrm{EX} = 2.5\,\si{\cm}$ (using only one \ac{LED}) up to $L_\mathrm{EX} = 29\,\si{\cm}$ (using all $12$ \acp{LED}). Since the switching probability of \ac{GFPD} depends on the energy supplied, increasing the number of \acp{LED} generally improves the erasure process. In this work, for simplicity, we focus on the case where all 12 \acp{LED} are used, i.e., $L_\mathrm{EX} = 29\,\si{\cm}$. The \ac{EX} is placed upstream next to the \ac{TX}.

Compared to the \ac{TX}, the use of the \ac{EX} is optional. Consequently, two scenarios are considered: one with and one without the use of the \ac{EX}. In the case without \ac{EX}, \ac{GFPD} can only switch back from the OFF state to the ON state by thermal relaxation. If we use the \ac{EX}, the \ac{EX} is turned on at the beginning of the experiment, i.e., the \ac{EX} is continuously active. Hence, the \ac{EX} enables the active reset of the \acp{GFPD} from the OFF state back to the ON state.  In consequence, with the \ac{EX} turned on, the number of ON state molecules available at the \ac{TX} is increased. In addition, the \ac{EX} effectively reduces memory effects in the system, i.e., it can be used to reduce the forms of \ac{ISI} specific to closed-loop \ac{MC} systems. We will discuss these forms of \ac{ISI} in detail in \Section{SubSec:ISI}. As a side effect, the \ac{EX} also contributes to photobleaching, cf. \Section{par:photobleaching}. Thus, the use of the \ac{EX} causes a tradeoff between \ac{ISI} mitigation and additional photobleaching.
\section{Physical Characterization of the Testbed}\label{sec:characterization}
In this section, we discuss the physical effects observed in the testbed, such as the involved fluid mechanics and different forms of \ac{ISI} present including those that occur exclusively in closed-loop systems.
\subsection{Signaling Molecule Propagation}\label{subsec:propagation}
In our testbed, the propagation of the \ac{GFPD} signaling molecules is mainly affected by advection, facilitated by the bulk background flow of the buffer liquid. In the following, we analyze the flow characteristics and its parameters in the testbed.

When a fluid of kinematic viscosity $\nu$ flows with average flow speed $v_{\mathrm{eff}}$ through a pipe of radius $r_\mathrm{T}$, either laminar flow or turbulent flow occurs. 
Which type of flow is dominant can be predicted by the Reynolds number $\text{Re} = \frac{2 r_\mathrm{T} \veff}{\nu}$ \cite[p. 14]{darby2017chemical}. For increasing Reynolds numbers, the transition from the laminar to the turbulent flow regime occurs around $\text{Re} \approx 2300$ \cite[p. 12]{schlichting2016boundary}. The viscosity of the buffer solution in our testbed equals the viscosity of water \cite{hink2000structural}, i.e., $\nu = 1.037\, \times 10^{-6} \,\si{ \meter \squared \per \second}$ (at $21\, \si{\celsius}$). 
Here, for the parameter regime in which the testbed is operated, we obtain $\text{Re} = 167$, i.e., laminar flow is dominant. This value of $\text{Re}$ corresponds to a medium-sized artery, where the Reynolds number is typically between 100 and 1000 \cite{caro2012mechanics}.
Furthermore, we determine the relative influence of diffusion on the fluid transport, which can be characterized by the dimensionless Péclet number $\text{Pe} = \frac{r_\mathrm{T} \veff}{\D}$ \cite[eq. (4.44)]{tabeling2023introduction}. Here, utilizing the diffusion coefficient $ \D = 1\, \times 10^{-10} \,\si{\meter\squared \per \second}$ of \ac{GFPD} \cite{Junghans2016DiffusionGFPD}, $\text{Pe} = 1.11\, \times 10^{6} \gg 1$ follows. Therefore, flow dominates over diffusion in the system, and the latter can be neglected as a result.
\subsection{ISI Caused by Closed-Loop Operation}\label{SubSec:ISI}
The closed-loop character of our testbed gives rise to four distinct forms of \ac{ISI}: \textit{channel \ac{ISI}}, \textit{inter-loop \ac{ISI}}, \textit{offset \ac{ISI}}, and \textit{permanent \ac{ISI}}. The latter three forms can only be observed in closed-loop systems. 
All forms of \ac{ISI} are discussed in detail in the following. 
\subsubsection{Channel ISI}\label{subsubsec:channel_isi}
Within the channel, the propagation of OFF state \ac{GFPD} molecules from \ac{TX} to \ac{RX} may result in the overlap of consecutive transmitted symbols, leading to \ac{ISI}.
The existence of channel \ac{ISI} has been demonstrated with existing \ac{MC} testbeds, e.g., \cite{grebenstein2018biological, wang2020understanding}.
Channel \ac{ISI} can be mitigated by introducing a guard interval of duration $T_\mathrm{G}$, during which the \ac{TX} is inactive, where the effectiveness of \ac{ISI} mitigation is contingent on the duration of the guard interval\footnote{While long guard intervals can theoretically compensate for the dispersive nature of the molecular channel, they are not practical in real-world scenarios, as increasing $T_\mathrm{G}$ also results in a direct reduction of the achievable data rate. Nevertheless, for illustration, consider the unrealistic case of using a guard interval of $T_\mathrm{G} = 6000\,\si{\second}$, which is one order of magnitude larger than the half-life of the thermal relaxation process, $T_{1/2} = 600\,\si{\second}$. In this case, we would not expect to observe channel, inter-loop, or offset \ac{ISI} as the long time interval between successive transmissions allows all \ac{GFPD} molecules in the OFF state to return to the ON state through thermal relaxation, even if the \ac{EX} is inactive.}.
\subsubsection{Inter-Loop ISI}\label{subsubsec:loop_isi}
Due to the closed-loop design of the testbed, inter-loop \ac{ISI} occurs, as the OFF state \ac{GFPD} molecules remain in the system and reenter the \ac{RX} after completing one loop. Hence, these molecules interfere not only with neighboring symbols (as is the case for channel \ac{ISI}) but also with symbols transmitted much later in time. In particular, OFF state \ac{GFPD} molecules, which are switched
to the OFF state by the \ac{TX}, may travel several times through the loop before they spontaneously switch back to the ON state. Thus, these signaling molecules are likely to affect the received signal multiple times, while the intensity, timing, and number of recurrences of inter-loop \ac{ISI} depend on the length of the loop and the half-life $T_{1/2}$ of \ac{GFPD}. The intensity of inter-loop \ac{ISI} can be reduced by using the \ac{EX}.
\subsubsection{Offset ISI}\label{subsubsec:offset_isi}
As \ac{GFPD} molecules in the OFF state disperse in the system over multiple loops, the inter-loop \ac{ISI}, characterized by distinct drops in fluorescence intensity, undergoes a gradual transformation to a temporal offset in the observed fluorescence intensity, which we refer to as offset \ac{ISI}. The intensity of offset \ac{ISI} depends on the irradiation intensity used at the \ac{TX}, symbol duration $\Ti$, and half-life $T_{1/2}$ of \ac{GFPD}. The intensity of offset \ac{ISI} can be reduced by using the \ac{EX}.
\subsubsection{Permanent ISI}\label{subsubsec:permanent_isi}
After prolonged exposure to light, \ac{GFPD} molecules may undergo an irreversible degradation as a consequence of photobleaching effects (cf. \Section{par:photobleaching}). While the decrease in fluorescence intensity resulting from offset \ac{ISI} is reversible, photobleaching is not. Thus, photobleaching causes an irreversible reduction of the number of functional \ac{GFPD} molecules available for modulation. Consequently, this leads to an overall decrease in the received fluorescence signal intensity over time. As a result, we observe a deterioration of the communication performance.
Since \ac{GFPD} molecules are irradiated by the \ac{EX}, the \ac{RX}, and the \ac{TX} in the testbed, all three light sources contribute to photobleaching. Therefore, we refer to the fractions caused by the \ac{TX}, \ac{EX}, and \ac{RX} as permanent \ac{ISI}, \ac{EX} bleaching, and measurement bleaching, respectively. Note that only photobleaching caused by the \ac{TX} is interpreted as \ac{ISI}, since it can be varied by varying the transmitted signal, whereas the photobleaching caused by \ac{EX} and \ac{RX} is independent of the transmitted sequence.
\section{Communication Scheme}\label{sec:communication}
This section introduces the proposed communication scheme, which includes modulation, synchronization, and detection.
\subsection{Modulation and Reception}\label{Sec:Modulation}
In \Sections{subsubsec:tx}{subsubsec:rx}, we discussed modulation and reception qualitatively, respectively. This section provides a formal description of the transmission parameters, their effects on the \ac{MC} signal, and the detection at the \ac{RX}.

To transmit a symbol $\symVar[k] \in \{0, 1, \ldots, \modOrder - 1\}$, the \ac{TX} is turned on with a corresponding light intensity $\txInten_\symVar$ for irradiation duration $\Ti$. Here, $\symIdx$ and $\modOrder = 2^\eta$ with $\eta \in \mathbb{N}$, where $\mathbb{N}$ is the set of positive integers, denote the symbol index and the modulation order, respectively. $\Ti$ is followed by the guard interval of duration $\Tg$, which results in $\Ts = \Ti + \Tg$, cf. \Section{subsubsec:tx}, and a data rate $R$ of
\begin{equation}\label{eq:data_rate}
    R = \frac{\log_2 \modOrder}{\Ts} \;.
\end{equation}
During the irradiation process, the \ac{GFPD} in the tube section of the \ac{TX} undergoes a state transition (from ON to OFF) when hit by the emitted photons, reducing the local fluorescence. Increasing light intensity $\txInten$ increases the hit probability, allowing control over the local fluorescence. This enables higher-order modulation with distinct fluorescence drops for different symbols $\symVar$. 
In particular, in this work, we modulate the light intensity ratio $\ratioInten_\symVar$ for symbol $\symVar$ as
\begin{equation}\label{eq:power_ratio}
    \ratioInten_\symVar = \frac{\txInten_\symVar}{\maxTxInten} = \frac{3 \symVar}{4(M-1)} + \frac{1}{4},
\end{equation}
i.e., $\ratioInten_\symVar$ is determined as the ratio of $\txInten_\symVar$ to the maximum available light intensity $\maxTxInten$. Eq.~\Equation{eq:power_ratio} shows that $\ratioInten_\symVar$ ranges from $1/4$ to $1$, i.e., we do not assign intensity ratio zero to any symbol. This guarantees that the transmission of all symbols is distinct from the idle channel state, i.e., a turned off \ac{TX}. This is necessary for symbol-by-symbol synchronization, as described in \Section{subsec:sync}.
On the \ac{RX} side, the spectrometer periodically measures the local fluorescence by sampling with time interval $\sampleInt$. After isolating the part of the signal close to the wavelength of interest, i.e., $\lambda_\mathrm{E} = 529 \, \si{\nm}$, and normalizing\footnote{The \textit{ThorSpectra software} internally scales the received fluorescence intensity in a non-transparent manner. Comparing absolute fluorescence values from different experiments is therefore not possible. Instead, it is useful to compare relative values, i.e., trends of the measured fluorescence intensities, which is guaranteed by the proposed normalization.} the maximum signal value to $1$, the discrete-time received signal $\recSig(\tn) \in [0,1]$ is obtained, where $\tn = n\sampleInt$, with $n \in \mathbb{N}_0$. Here, $\mathbb{N}_0$ denotes the set of non-negative integers.
\subsection{Synchronization}\label{subsec:sync}
Synchronization is an integral part of any communication system and is required for demodulation. While in stationary environments it can be sufficient to synchronize once at the beginning of the transmission of a data packet, the task of synchronization becomes critical -- and more challenging -- when dealing with non-stationary communication channels. Sources of this non-stationarity can be, for example, closed-loop \ac{ISI} effects, \ac{TX} and \ac{RX} movement, or a time-varying flow velocity. In this section, we propose synchronization schemes consisting of a transmission start detection and symbol-by-symbol synchronization. Hence, they address the aforementioned synchronization challenges.
\subsubsection{Transmission Start Detection}\label{subsec:transmission}
Prior to the initiation of demodulation, the \ac{RX} must detect that data transmission has started. To this end, a simple threshold-based trigger scheme is used, where transmission is assumed to have started when $\recSig(\tn)$ drops below a threshold $\trainThresh$. For this, $\recSig(\tn)$, which prior the to first transmission contains only noise, is used to estimate the noise statistics. Then, $\trainThresh$ is determined based on the noise statistic. We assume that the noise arises from many small, independent sources of randomness, e.g., thermal noise in the spectrometer, photon noise, i.e., the distribution of photons emitted by coherent light \cite{mandel1959fluctuations} from the \ac{RX} \ac{LED}, etc. Hence, applying the central limit theorem, we model the noise as Gaussian distributed.

Formally, before transmission has started, we collect a set of samples $\trainSet[\trainIdx] \triangleq \{\recSig(\tn)|\n \in \{\trainIdx \trainLength, \trainIdx \trainLength + 1, \hdots, (\trainIdx+1) \trainLength -1 \}\}$ of size $\trainLength$ for $\trainIdx \in \mathbb{N}_0$. Next, $\trainSet[\trainIdx]$ is used to fit a Normal distribution with mean $\trainMean[\trainIdx] = \frac{1}{\trainLength}\sum_{\recSig(\tn) \in \trainSet[\trainIdx]}\recSig(\tn)$ and variance $\trainVariance[\trainIdx] = \frac{1}{\trainLength-1}\sum_{\recSig(\tn) \in \trainSet[\trainIdx]} (\recSig(\tn) - \trainMean[\trainIdx])^2$ assuming independent and identically distributed samples, i.e., $\recSig(\tn)\,\sim\,\mathcal{N}(\trainMean[\trainIdx], \trainVariance[\trainIdx])$. Finally, we obtain the threshold value $\trainThresh[\trainIdx] = \Phi^{-1}(\pFA; \trainMean[\trainIdx], \trainVariance[\trainIdx])$ for which a desired false alarm probability $\pFA$, i.e., the residual risk of an incorrect decision that transmission has started, is achieved. Here, $\Phi^{-1}(\cdot)$ denotes the inverse of the Gaussian cumulative distribution function.

The obtained $\trainThresh[\trainIdx]$ is applied to the next sample set $\trainSet[\trainIdx+1]$. The transmission start is detected based on
\begin{equation}
    \tts = \min_{\forall \trainIdx}\{\tn|\recSig(\tn) \leq \trainThresh[\trainIdx],\, \recSig(\tn) \in \trainSet[\trainIdx+1]\}\;,
\end{equation}
where $\tts$ denotes the time at which transmission has started. Since $\recSig(\tn)$ may decreases (slowly) over time due to photobleaching, this process is continuously repeated to ensure that $\trainThresh$ is periodically adjusted until a transmission start has been detected\footnote{Note that detecting the end of transmission is not needed, as we assume fixed-length messages.}.
\subsubsection{Symbol Synchronization}\label{sec:symbol_sync}
Synchronization guarantees a temporal alignment of the \ac{RX} with the \ac{TX}. For the testbed, we use a symbol-by-symbol synchronization approach \cite{jamali2017symbol}, which correlates the processed received signal\footnote{In particular, we use $\Tilde{\recSig}(\tn) \in \{1 - \recSig(\tn), \recSig(t_{\n+1}) - \recSig(\tn)\}$, as explained in detail in the next paragraph.} $\Tilde{\recSig}(\tn)$ with a receive template filter $g(\tn)$ to obtain the synchronization metric
\begin{equation}
  \syncMetric (\tn) = \vec{\filter}^\top \Tilde{\vec{\recSig}} (\tn) \;,
\label{sync_metric}
\end{equation}
based on which the symbol start time $\tsEst[\symIdx]$ is estimated\footnote{Note that $\tsEst[k]$ does not correspond to the actual symbol start time, as the channel-related propagation delay is not known at the \ac{RX}. Instead, it represents the \ac{RX}'s estimate of when a new symbol starts based on the selected template signal/filter.}.
Here, $\vec{\filter} = \big[\, g(t_{0}) \quad g(t_{1}) \quad \hdots \quad g(t_{\filterLength-1})\,\big]^\top$ and $\Tilde{\vec{\recSig}}(\tn) = \big[\,\Tilde{\recSig}(\tn) \quad \Tilde{\recSig}(t_{n+1}) \quad \hdots \quad \Tilde{\recSig}(t_{n+\filterLength-1})\,\big]^\top$ are the vector representations of $g(\tn)$ and $\Tilde{\recSig}(\tn)$, respectively, where $\filterLength$ is the filter length and $[\cdot]^\top$ denotes the transpose operator.
Two synchronization schemes are employed: \Ac{CS} and \ac{DCS}.
\begin{itemize}
    \item \textit{Correlation-based Synchronization (CS)}: The \ac{CS} scheme utilizes the cross-correlation between the processed received signal $\Tilde{\recSig}(\tn) = 1 - \recSig(\tn)$, which captures the deviation from maximum fluorescence, and a corresponding receive filter $g(\tn) \in \{g^{\mathrm{D}}_{\mathrm{C}}(\tn), g^{\mathrm{B}}_{\mathrm{C}}(\tn)\}$. Details of how the data-based receive filter $g^{\mathrm{D}}_{\mathrm{C}}(\tn)$, referred to as \ac{SCF}, and the blind receive filter $g^{\mathrm{B}}_{\mathrm{C}}(\tn)$, referred to as \ac{BCF}, are obtained, respectively, are provided in \Section{matched_filter}.
    \item \textit{Differential Correlation-based Synchronization (DCS)}: The \ac{DCS} scheme is designed to mitigate the effects of slowly time-varying processes, such as offset-\ac{ISI}, and utilizes the cross-correlation between the \textit{differential} received signal $\Tilde{\recSig}(\tn) = \diffRecSig(\tn) = \recSig(t_{\n+1}) - \recSig(\tn)$ and a corresponding \textit{differential} receive filter $g(\tn) \in \{g^{\mathrm{D}}_{\mathrm{D}}(\tn), g^{\mathrm{B}}_{\mathrm{D}}(\tn)\}$. Note that in this case the corresponding vectors $\vec{\filter}$ and $\Tilde{\vec{\recSig}} (\tn)$ in \Equation{sync_metric} have dimension $\filterLength-1$, as the forward difference cannot be computed for the last vector entry. Details of how the data-based differential receive filter $g^{\mathrm{D}}_{\mathrm{D}}(\tn)$, referred to as \ac{SDCF}, and the blind differential receive filter $g^{\mathrm{B}}_{\mathrm{D}}(\tn)$, referred to as \ac{BDCF}, are obtained, respectively, are provided in \Section{matched_filter}.
  \end{itemize}
\paragraph{Symbol-by-symbol synchronization}
In the proposed symbol-by-symbol scheme, synchronization for the symbol in symbol interval $k$ relies on the last estimated symbol start time $\tsEst[k-1]$. A good initial guess for the start of the current symbol is given by $\tsInit[k] = \tsEst[k-1] + \Ts$. Note that $\tsInit[k]$ would also be the best estimate, if the previous estimate was accurate and the channel had not changed. Next, we define a search interval around $\tsInit[k]$ as $\searchInt[k] \triangleq [\tsInit[k] - \searchRadius \Ts, \tsInit[k] + \searchRadius \Ts]$. Here, $\searchRadius \in [0,1]$ denotes the normalized search radius, which can be adjusted based on the channel statistics, i.e., the stationarity of the channel.
The start time $\tsEst[k]$ of the current symbol interval $k$ is obtained as
\begin{equation}
  \tsEst[k] = \argmax_{\tn \in \searchInt[\symIdx]} \syncMetric(\tn)\;.
  \label{eq:estimation_symbol_start}
\end{equation}
This process is carried out for each symbol interval until the end of the transmission. The proposed synchronization method can adjust to fluctuations of the propagation delays in the channel within a single symbol interval as long as the true symbol start time is within the search interval.
\paragraph{Synchronization initialization}
Since for the first symbol interval there is no previous estimate $\tsEst[-1]$, we utilize the detected transmission start time as initial estimate, $\tsInit[0] = \tts$, and set the search radius to $\searchRadius=0.5$ -- resulting in search interval $\searchInt[0] = [\tsInit[0] - \frac{\Ts}{2}, \tsInit[0] + \frac{\Ts}{2}]$. Even if $\tts$ does not correspond to a false alarm, it may still be a suboptimal estimate for the first symbol start. Therefore, choosing search radius $\searchRadius=0.5$ ensures that the best estimate for the first symbol start time is within the search interval. Synchronization then proceeds as described above.
\subsection{Receive Filters} \label{matched_filter}
\begin{figure*}[!tbp]
    \centering
    \begin{subfigure}[b]{0.49\textwidth}
        \caption{}
        \includegraphics[width=\textwidth]{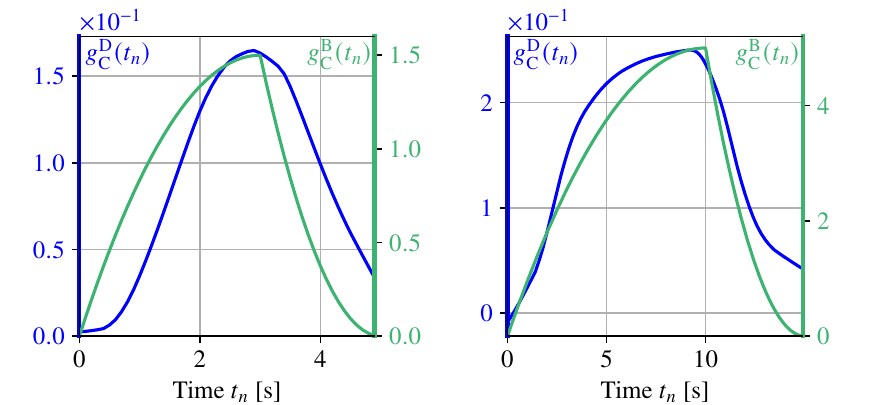}
        \label{fig:BlindCorrFilter}
    \end{subfigure}
    \begin{subfigure}[b]{0.49\textwidth}
        \caption{}
        \includegraphics[width=\textwidth]{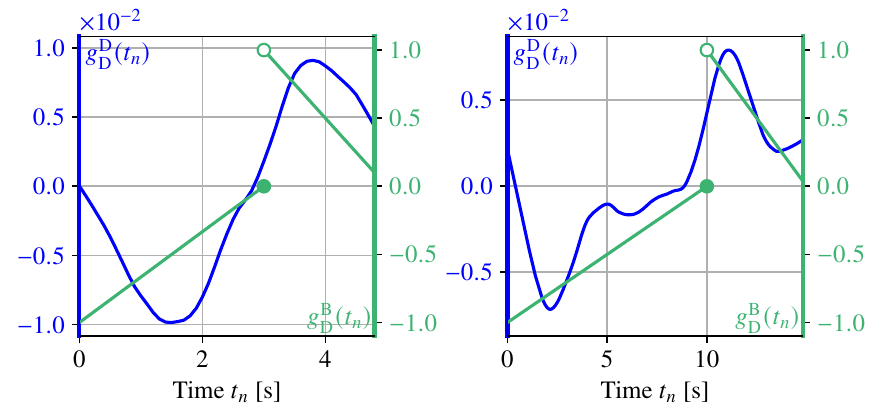}
        \label{fig:BlindDiffFilter}
    \end{subfigure}
    \begin{subfigure}[b]{\textwidth}
        \includegraphics[width=\textwidth]{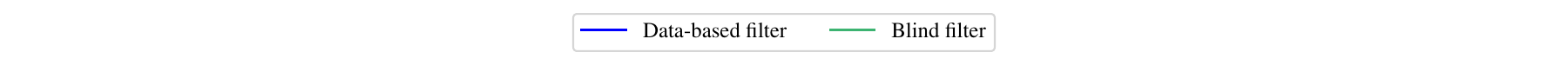}
    \end{subfigure}
    \caption{Data-based receive filters and blind receive filters in comparison for two different configurations: $\Ti=\SI{3}{\second}$, $\Ts=\SI{5}{\second}$ (on the left) and $\Ti=\SI{10}{\second}$, $\Ts=\SI{15}{\second}$ (on the right) for the \acs{CS} (a) and the \acs{DCS} scheme (b), respectively.}
    \label{fig:ReceiveFilter}
\end{figure*}
Two different approaches are used to obtain receive filters for the \ac{CS} and the \ac{DCS} schemes, namely a data-based approach and a blind filter approach. Although these filters are akin to a matched filter, we do no refer to them as such, as we cannot guarantee their optimality \ac{wrt} the achieved signal-to-noise ratio.
\subsubsection{Data-Based Receive Filters}\label{data_filter}
For each $\Ti$ considered in this work, a separate experiment was performed to obtain the typical shape of the received signal for that irradiation duration, on the basis of which the data-based filter was determined. Each of these experiments consisted of a binary transmission of a single bit 1 repeated 11 times at 60 second intervals. The 11 individual received signals obtained in this way are referred to as \acp{SR} $\SR(\tn)$. To derive the data-based filter, the first \ac{SR} is discarded and the remaining \acp{SR} are averaged and further smoothed using locally weighted polynomial regression \cite{cleveland1988locally} to obtain $\avgSR(\tn)$. Ignoring the first \ac{SR} is necessary because in the beginning almost all \acp{GFPD} are in the equilibrium state, cf. \Section{subsec:GFPD}, and therefore the fluorescence decay for the first \ac{SR} is higher than for the following \acp{SR}.
Then, the \ac{SCF}, $g^{\mathrm{D}}_{\mathrm{C}}(\tn)  = 1 - \avgSR(\tn)$, $n \in \{0, 1, \ldots, \filterLength-1\}$, and the \ac{SDCF}, $g^{\mathrm{D}}_{\mathrm{D}}(\tn)  = \avgSR(t_{\n+1}) - \avgSR(\tn)$, $n \in \{0, 1, \ldots, \filterLength-2\}$, are obtained by limiting them to an $\filterLength$ and $\filterLength-1$ samples long signal, respectively, i.e., limiting them such that their lengths match the lengths of the corresponding $\Tilde{\vec{\recSig}}(\tn)$, cf. \Equation{sync_metric}. The resulting receive filters $g^{\mathrm{D}}_{\mathrm{C}}(\tn)$ and $g^{\mathrm{D}}_{\mathrm{D}}(\tn)$ are shown in blue in \Figure{fig:BlindCorrFilter} and \Figure{fig:BlindDiffFilter}, respectively, exemplarily for two different symbol durations $\Ts$. \Figure{fig:BlindCorrFilter} shows that $g^{\mathrm{D}}_{\mathrm{C}}(\tn)$ increases over the irradiation durations $\Ti=\SI{3}{\second}$ (left subplot in \Figure{fig:BlindCorrFilter}) and $\Ti=\SI{10}{\second}$ (right subplot in \Figure{fig:BlindCorrFilter}), and decreases during the subsequent guard interval. For $\Ts=\SI{5}{\second}$ (left subplot in \Figure{fig:BlindCorrFilter}), $g^{\mathrm{D}}_{\mathrm{C}}(\tn)$ is bell-shaped, while for the longer symbol duration, $\Ts=\SI{15}{\second}$ (right subplot in \Figure{fig:BlindCorrFilter}), we see that $g^{\mathrm{D}}_{\mathrm{C}}(\tn)$ reaches a value of $0.25$, before decreasing after $\tn = \Ti = \SI{10}{\second}$. In \Figure{fig:BlindDiffFilter}, we see that the shape of $g^{\mathrm{D}}_{\mathrm{D}}(\tn)$ follows the derivative of $- g^{\mathrm{D}}_{\mathrm{C}}(\tn)$. Hence, it has a negative and a positive dip. For $\Ts=\SI{15}{\second}$ (right subplot in \Figure{fig:BlindDiffFilter}), $g^{\mathrm{D}}_{\mathrm{D}}(\tn)$ shows some fluctuations around $\tn=\SI{6}{\second}$, caused by noise that is amplified by the differentiation operation.
\subsubsection{Blind Receive Filter}\label{blind_filter}
In cases where the data-based filter cannot be obtained in advance, we propose blind filters as surrogates for both the \ac{CS} and \ac{DCS} schemes, respectively. For these blind filters, we aim to capture roughly the shape of the data-based filters, with the goal of achieving a solution that is applicable across a broad range of unknown parameters without significant performance loss. For the sake of a simple and practical implementation, we limit the modeling of the blind filters to piecewise polynomials. Hence, the impulse responses of the proposed blind filters \ac{BCF}, $g^{\mathrm{B}}_{\mathrm{C}}(\tn)$, and \ac{BDCF}, $g^{\mathrm{B}}_{\mathrm{D}}(\tn)$, are given as follows,
\begin{equation}\label{eq:blind_corr_filter}
\blindCorrFilter(\tn) =
\begin{cases}
     - \frac{\tn^2}{2\Ti} + \tn, & 0\leq \tn \leq \Ti\\
     \frac{0.5\Ti}{(\Ts-\Ti)^2}(\tn^2 -2\Ts \tn + \Ts^2), &  \Ti < \tn \leq  \Ts\\
     0, & \text{otherwise}
\end{cases}
\end{equation}
and
\begin{equation}\label{eq:blind_diff_corr_filter}
\blindDiffCorrFilter(\tn) =
    \begin{cases}
        \frac{\tn}{\Ti} - 1, & 0\leq \tn \leq \Ti\\
        \frac{\Ts - \tn}{\Ts-\Ti}, &  \Ti < \tn \leq  \Ts\\
        0, & \mathrm{otherwise}
    \end{cases}\,,
\end{equation}
respectively. $g^{\mathrm{B}}_{\mathrm{C}}(\tn)$ and $g^{\mathrm{B}}_{\mathrm{D}}(\tn)$ are plotted in green in \Figure{fig:BlindCorrFilter} and \Figure{fig:BlindDiffFilter}, respectively, exemplarily for two different symbol durations $\Ts$. \Figure{fig:BlindCorrFilter} shows that the shapes of the data-based filter $g^{\mathrm{D}}_{\mathrm{C}}(\tn)$ and the blind filter $g^{\mathrm{B}}_{\mathrm{C}}(\tn)$ are quite similar, while those of the differential filters $g^{\mathrm{D}}_{\mathrm{D}}(\tn)$ and $g^{\mathrm{B}}_{\mathrm{D}}(\tn)$ in \Figure{fig:BlindDiffFilter} are more different. We will show in \Section{sssec:effect_filter_and_detection} that despite these visible differences, the use of blind filters only leads to a small loss in performance compared to the use of data-based filters. In addition, blind filters do not require data collection, which justifies their use. Note that the blind filters are defined independent of each other as this leads to better results, i.e., $g^{\mathrm{B}}_{\mathrm{D}}(\tn)$ is not necessarily the sectional derivative of $g^{\mathrm{B}}_{\mathrm{C}}(\tn)$. Furthermore, note that the scaling of the data-based filters and the blind filters are different. However, for a given synchronization scheme, adjusting the scaling of any receive filter has no impact on the decisions derived from \Equation{sync_metric} and \Equation{eq:estimation_symbol_start}, since the operation in \Equation{sync_metric} is linear. Therefore, no scaling adjustment is needed.
\subsection{Symbol Detection}\label{subsec:symbol_detection}
For symbol detection, we use a single detection sample $\detecSample[\symIdx]$. Here, the detection sample is chosen to be the filtered received signal for the estimated symbol start time $\tsEst[\symIdx]$, which was determined by the proposed synchronization scheme, i.e., 
\begin{equation}
\detecSample[\symIdx]=\vec{\filter}^\top \Tilde{\vec{\recSig}} (\tsEst[\symIdx]) \;. \label{eq:detection_sample}
\end{equation}
In \Equation{eq:detection_sample}, correlation-based detection and differential correlation-based detection are employed, if \ac{CS} and \ac{DCS} was used for synchronization, respectively. Hence, when \ac{CS} and \ac{DCS} are mentioned in the following, this therefore also refers to the detection scheme used.

For detection, we employ an adaptive (multi-)threshold detector that periodically adjusts its thresholds based on previously detected symbols to account for the gradual reduction of the fluorescence over time due to photobleaching. Let $\detecThreshSet[\detecThreshSetIdx] \triangleq \{\detecThresh_0, \detecThresh_1, \hdots, \detecThresh_{\modOrder-2}\}$ denote the set of detection thresholds, where $\detecThresh_{\detecThreshIdx}$, $\detecThreshIdx \in \{0, 1, \hdots, \modOrder-2\}$, and $\detecThreshSetIdx$ denote the individual threshold and the threshold set index, respectively. Additionally, we assume \ac{wlog} that $\detecThreshSet[\detecThreshSetIdx]$ is ordered in ascending order, i.e., $\detecThresh_\detecThreshIdx < \detecThresh_{\detecThreshIdx+1}$. For the employed single-sample detection, the transmitted symbol $\estSym[\symIdx]$ is determined as follows
\begin{equation}
\estSym[\symIdx] = \max\{\detecThreshIdx | \detecThreshIdx \in \{0, 1, \hdots, \modOrder-2\} \wedge \detecSample[\symIdx] \geq \detecThresh_\detecThreshIdx\} \;, \label{eq:estimated_symbol}
\end{equation}
i.e., the symbol decision is based on $\detecSample[\symIdx]$ and $\detecThreshSet[\detecThreshSetIdx]$ only. Details on how $\detecThreshSet[\detecThreshSetIdx]$ is determined are provided next.
\subsubsection{Detection Initialization}\label{sec:init_thresholds}
To determine the initial set of thresholds, $\detecThreshSet[0]$, each transmission starts with a random sequence of $\nSkip$ symbols.
This sequence is used to allow the system to settle\footnote{The experiment is initially in a settling phase caused by the offset \ac{ISI}, which develops slowly over successive transmissions and remains relatively constant after several symbols. \Section{sssec:effect_modulation_order} presents experimental results that illustrate the transient phase of the testbed.} and is not considered for detection. Next, $\nPilots$ pilot symbols are transmitted, where each $\symVar \in \{0, 1, \ldots, \modOrder - 1\}$ is sent at least once. The corresponding detection samples $\detecSample[k]$ form the initial sets $\symbolSampleSet_{\symVar}[\detecThreshSetIdx = 0] \triangleq \{d[\mu]|\mu \in \{\nSkip, \nSkip+1, \ldots, \nSkip+\nPilots-1\} \wedge \symVar[\mu]=i\}$, $\symVar \in \{0, 1, \ldots, \modOrder - 1\}$. The average of each set can be computed as follows
\begin{equation}
    \Bar{S}_\symVar[\detecThreshSetIdx] = \frac{1}{|\symbolSampleSet_{\symVar}[\detecThreshSetIdx]|}\sum_{s\in \symbolSampleSet_{\symVar}[\detecThreshSetIdx]} s \;.
    \label{eq:average_set}
\end{equation}
Here, $|\cdot|$ denotes the cardinality of a set. Hence, we use $\detecThreshSetIdx = 0$ in \Equation{eq:average_set} to determine the averages $\Bar{S}_\symVar[\detecThreshSetIdx = 0]$ of the initial sets $\symbolSampleSet_{\symVar}[\detecThreshSetIdx = 0]$. Finally, the threshold values are computed as the means of the $\Bar{S}_\symVar[l]$ of adjacent sets:
\begin{equation}\label{eq:nt}
    \detecThresh_\detecThreshIdx[\detecThreshSetIdx] = \frac{\avgSymbolSampleSet_{\detecThreshIdx+1}[\detecThreshSetIdx] + \avgSymbolSampleSet_{\detecThreshIdx}[\detecThreshSetIdx]}{2}\,,
\end{equation}
for $\detecThreshIdx \in \{0, 1, \hdots, \modOrder-2\}$. Thus, to obtain the initial set of thresholds, $\detecThreshSet[0]$, we use $\detecThreshSetIdx = 0$ in \Equation{eq:nt}.
\subsubsection{Adaptation Algorithm}\label{sec:adaption_algorithm}
The duration for which the threshold set is valid has an upper limit that is determined by the coherence time of the channel. Therefore, a new set of thresholds must be determined after a certain number of detected symbols $\nCoherence$. In contrast to the initial set of thresholds $\detecThreshSet[0]$, which is determined based on the $\nPilots$ pilot symbols, the \textit{update} of the set of thresholds requires only previously detected \textit{data symbols}, i.e., no pilot symbols are used to update the threshold values. The corresponding update algorithm is described next.

For this, we relate the symbol index $\symIdx$ to the currently valid threshold set index $\detecThreshSetIdx$ via $\detecThreshSetIdx = \lfloor \frac{\symIdx-\nSkip-\nPilots}{\nCoherence}\rfloor$, where $\lfloor \cdot \rfloor$ denotes the floor function. Note that $\detecThreshSetIdx$ is not defined for $\symIdx < \nPilots + \nSkip$ because the first thresholds are computed only after the $\nPilots$ pilot symbols have been received.

Similar to the initial sets $\symbolSampleSet_\symVar[\detecThreshSetIdx = 0]$, we define sets
\begin{equation}
    \symbolSampleSet_\symVar[\detecThreshSetIdx]\triangleq\{\detecSample[\mu]|\mu \in \{\detecThreshSetIdx\nCoherence+\nSkip+\nPilots-\nWindow,..., \detecThreshSetIdx\nCoherence+\nSkip+\nPilots-1\} \wedge \estSym[\mu]=\symVar\}\;,
    \label{eq:sampleSets}
\end{equation}
for $\detecThreshSetIdx>0$ and $\nCoherence+\nPilots-\nWindow \geq 0$. Here, $\nWindow$ denotes the window width that determines the number of past samples $\detecSample[k]$ used for reevaluation. Finally, using \Equation{eq:average_set} and \Equation{eq:nt}, we obtain the new set of thresholds $\detecThreshSet[\detecThreshSetIdx]$.

In contrast to the pilot symbol sequence, now it is not guaranteed that all symbol values will be present in the $\nWindow$ previously received symbols, i.e., some sets may remain empty ($|\symbolSampleSet_\symVar[\detecThreshSetIdx]| = 0$). For this special case, the corresponding new thresholds cannot be determined directly using \Equation{eq:average_set} and \Equation{eq:nt}. We determine the differences between the new thresholds and the previous thresholds for the subset of thresholds for which this is possible, i.e., $\threshDiff_q[\detecThreshSetIdx] =\detecThresh_q[\detecThreshSetIdx] - \detecThresh_q[\detecThreshSetIdx-1]$ for $q \in \mathcal{Q}\triangleq\{\detecThreshIdx \,|\,|\symbolSampleSet_{\detecThreshIdx+1}[\detecThreshSetIdx]|\neq0 \wedge |\symbolSampleSet_{\detecThreshIdx}[\detecThreshSetIdx]|\neq 0\}$ with $\detecThreshIdx \in \{0, 1, \hdots, \modOrder-2\}$. The thresholds corresponding to the empty sets are subsequently obtained by adjusting the previous threshold value according to the \textit{average} change across all thresholds $\avgThreshDiff[\detecThreshSetIdx] = \frac{1}{|\mathcal{Q}|}\sum_{q \in \mathcal{Q}}\threshDiff_q[\detecThreshSetIdx]$ as $\detecThresh_\detecThreshIdx[\detecThreshSetIdx] = \detecThresh_\detecThreshIdx[\detecThreshSetIdx-1] + \avgThreshDiff[\detecThreshSetIdx]$ for $\detecThreshIdx \notin \mathcal{Q}$. For the rare case, for which no new threshold can be determined, i.e., $|\symbolSampleSet_{\detecThreshIdx}[\detecThreshSetIdx]|=0 \,\vee\, |\symbolSampleSet_{\detecThreshIdx+1}[\detecThreshSetIdx]|=0,\,\forall \detecThreshIdx$, the thresholds retain their previous values, i.e., $\detecThreshSet[\detecThreshSetIdx]=\detecThreshSet[\detecThreshSetIdx-1]$.
\section{Performance Metrics}\label{sec:metric}
In this section, we describe the quantitative and qualitative communication-specific metrics employed to evaluate \ac{MC} performance in the proposed testbed.
\subsection{Bit Error Rate and Data Rate}\label{sec:ber_and_data_rate}
We define an \textit{empirical} \ac{BER} for our system. The \ac{BER} is a standard metric for quantifying the performance of any communication scheme. For the bit-to-symbol mapping, we use Gray mapping \cite[p. 100]{proakis2008digital} $G: \{0,1\}^{\log_{2}(M)\nSymb} \rightarrow \{0, \ldots, M - 1\}^{\nSymb}$. Hence, the bit sequences corresponding to symbol sequences $\mathbf{i} \in \{0, 1, \ldots , M - 1\}^{\nSymb}$ and $\hat{\mathbf{i}} \in \{0, 1, \ldots , M - 1\}^{\nSymb}$, i.e., the transmitted and estimated symbol sequences, are obtained as $\mathbf{i}_{G} = G^{-1}(\mathbf{i})$ and $\hat{\mathbf{i}}_G = G^{-1}(\hat{\mathbf{i}})$, respectively. In our case, since the use of pilot symbols is required for calibration of all our detection techniques, the \ac{BER} is obtained as
\begin{equation}
    \mathrm{BER} = \frac{\sum_{b = \log_2(M)\nPilots}^{\log_{2}(M)\nSymb - 1} \left|\mathbf{i}_{G}[b] - \hat{\mathbf{i}}_G[b]\right|}{(\nSymb - \nPilots)\log_2(M)}\;.
    \label{eq:BER}
\end{equation}
Note that this equals the Hamming distance between $\mathbf{i}_{G}$ and $\hat{\mathbf{i}}_G$ normalized to the bit sequence length excluding the sequence part corresponding to the pilot symbols $\nPilots$. Thus, the empirically defined \ac{BER} is the number of erroneously demodulated bits scaled by the number of all transmitted bits.
\subsection{Absolute Mean Euclidean Distance}
In some cases the \ac{BER} is not insightful, e.g., if the transmission sequence length is too short for an accurate estimation of the \ac{BER} or if the \ac{MC} system is in the low \ac{BER} regime, where irrespectively of the chosen settings, for the evaluated transmission sequence length, $\textnormal{BER}=0$ is obtained.
For such cases, we introduce the \ac{AMED} metric as follows
\begin{equation}
   \AMED = \min_{i' \neq i''} |\Bar{S}_{i'}[l] - \Bar{S}_{i''}[l]| \;; \qquad \forall i', i'' \in \{0, 1, \ldots, \modOrder - 1\} \;,
    \label{eq:amed}
\end{equation}
where we employ \Equation{eq:average_set} and \Equation{eq:sampleSets}.

As can be seen from \eqref{eq:amed}, the \ac{AMED} is the minimum Euclidean distance between the means of the samples $\detecSample[\symIdx]$ received over $\nWindow$ symbol intervals. Consequently, the \ac{AMED} reflects the extent to which the received samples of the various symbols differ. Compared to the \ac{BER}, the \ac{AMED} provides reliable values even for small numbers of samples, i.e., small $\nWindow$, and can therefore serve as an initial indicator of performance for cases where a \ac{BER} evaluation is not meaningful or not possible.

\subsection{Eye Diagram}
Eye diagrams are a well-established qualitative method to evaluate a communication system's performance \cite[p. 603]{proakis2008digital}. The use of eye diagrams to evaluate \ac{MC} systems has been introduced in \cite{farsad2017novel}. Plotting the received signal modulo the symbol duration generates the eye pattern. Effects such as \ac{ISI}, various system parameters, including the use of a guard interval, and the effects of varying the modulation order affect the eye pattern and the eye opening, which can provide insights for system design and performance.

\section{Experimental Results and Communication Performance Evaluation}\label{sec:results}
In this section, we evaluate our testbed based on the previously proposed metrics and determine the impact of different system parameters. To obtain these results, in total more than 250 kbit of data were transmitted via \ac{MC} in the testbed.
\subsection{Illustration of the Different Types of ISI}\label{subsubsec:concret_char_ISI}
\begin{figure*}[!tbp]
\centering
  \includegraphics[width=\textwidth]{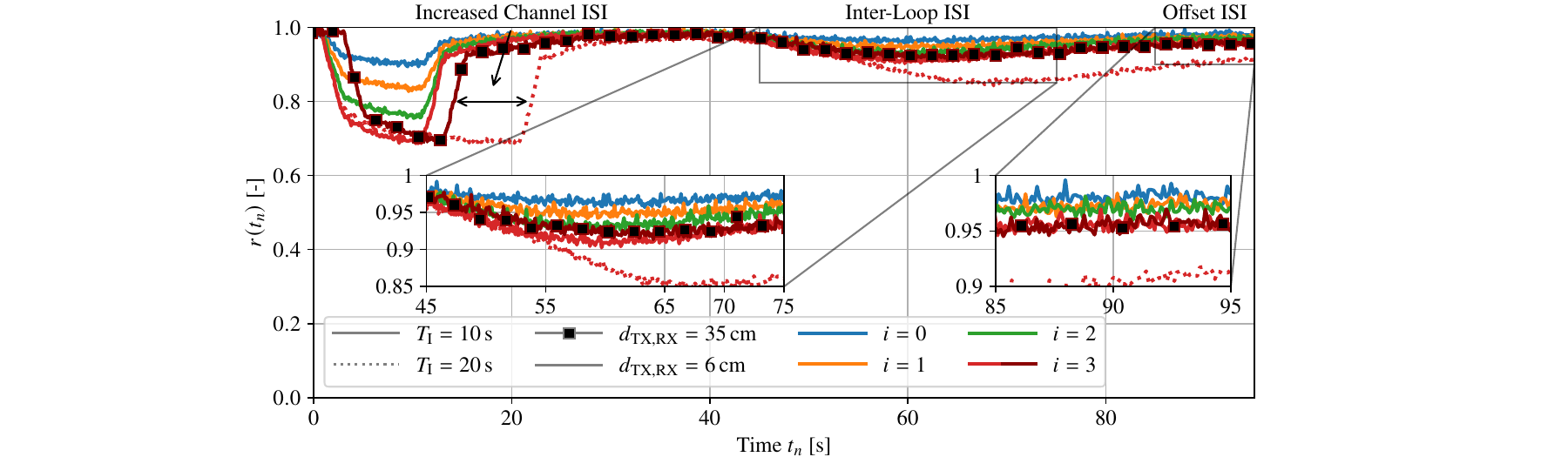}
  \caption{$\recSig(\tn)$ of single symbol transmissions with $\symVar \in \{0, 1, 2, 3\}$ for different irradiation durations $\Ti = 10 \,\si{\second}$ and $\Ti = 20 \,\si{\second}$, and \ac{TX}-\ac{RX} distances $d_{\mathrm{TX},\mathrm{RX}} = 6 \,\si{\centi\meter}$ and $d_{\mathrm{TX},\mathrm{RX}} = 35 \,\si{\centi\meter}$. Three forms of \ac{ISI} are visualized: channel \ac{ISI}, inter-loop \ac{ISI}, and offset \ac{ISI}.}
  \label{fig:SIR}
\end{figure*}
The following experiments were conducted to illustrate three different types of \ac{ISI} that occur in our testbed. The results are summarized in \Figure{fig:SIR}. In particular, individual symbols $\symVar \in \{0, 1, 2, 3\}$, indicated by different colors, are sent. Furthermore, for $\symVar = 3$, we show results for two illumination durations, $\Ti= 10 \,\si{\second}$ (solid line) and $\Ti = 20 \,\si{\second}$ (dotted line), and two different transmission distances, $d_{\mathrm{TX},\mathrm{RX}} = 6 \, \si{\centi\meter}$ (no marker) and $d_{\mathrm{TX},\mathrm{RX}} = 35 \, \si{\centi\meter}$ (rectangular markers). \Figure{fig:SIR} shows the corresponding received signals $\recSig(\tn)$. In these experiments, the \ac{EX} was turned off to increase the visibility of the \ac{ISI} effects.

For $\Ti= 10 \,\si{\second}$, \Figure{fig:SIR} shows two drops in $\recSig(\tn)$. The first drop around $10\, \si{\second}$ corresponds to the desired pulse sent by the \ac{TX}, after which the signal returns back to its original level at around $30\, \si{\second}$. The second drop, visible around $60\, \si{\second}$, corresponds to inter-loop \ac{ISI}, while the offset \ac{ISI} is visible in the inset at $90 \, \si{\second}$. The time difference of approximately $50\, \si{\second}$ between the two minima confirms the order of magnitude of the estimated loop time of $T_\mathrm{L} = 57\,\si{\second}$, cf. \Section{subsec:structure}.
Both the inter-loop \ac{ISI} and the offset \ac{ISI} increase with increasing $\symVar$.
Since permanent \ac{ISI} caused by photobleaching occurs after many transmission cycles only, it is not visible in \Figure{fig:SIR}.
We observe from \Figure{fig:SIR} that $\recSig(\tn)$ returns back close to its initial value $\recSig(\tn) = 1.0$ at a later time when the irradiation time is increased from $\Ti = 10 \, \si{\second}$ (solid lines) to $\Ti = 20 \, \si{\second}$ (dotted line), consequently increasing the channel \ac{ISI}. Furthermore, we observe that the depth of the fluorescence intensity drop of the first pulse remains the same, while the inter-loop \ac{ISI} intensity increases significantly. From this, we infer that both the irradiation time $\Ti$ and which symbol $\symVar$ is transmitted have an influence on the inter-loop \ac{ISI}.
Moreover, \Figure{fig:SIR} shows that the irradiation time $\Ti$ and which symbol $\symVar$ is transmitted also have an impact on the offset \ac{ISI}.
These observations suggest that there exists an optimal irradiation duration $\Ti$. In particular, there exists a trade-off as $\Ti$ should be chosen large enough to produce a detectable signal, but small enough to prevent the undesired accumulation of \ac{ISI}, as all forms of \ac{ISI} increase when $\Ti$ is increased.

Finally, we observe from \Figure{fig:SIR} that increasing the transmission distance from $d_{\mathrm{TX},\mathrm{RX}} = 6 \, \si{\centi\meter}$ (no marker) to $d_{\mathrm{TX},\mathrm{RX}} = 35 \, \si{\centi\meter}$ (rectangular markers) results in a later arrival of the first pulse at the \ac{RX}, which is expected as the signal molecules have to propagate a longer distance. Furthermore, we observe that the pulse shape remains the same. Hence, we conclude that even a transmission distance of $d_{\mathrm{TX},\mathrm{RX}} = 35 \, \si{\centi\meter}$ is too small for diffusion to have an impact.
\subsection{Transmission and Error-Free Detection of 90 kbit}\label{sec:error_free_80kbit}
\begin{figure*}[!tbp]
    \centering

    \begin{subfigure}[b]{1\textwidth}
        \caption{}
        \includegraphics[width=\textwidth]{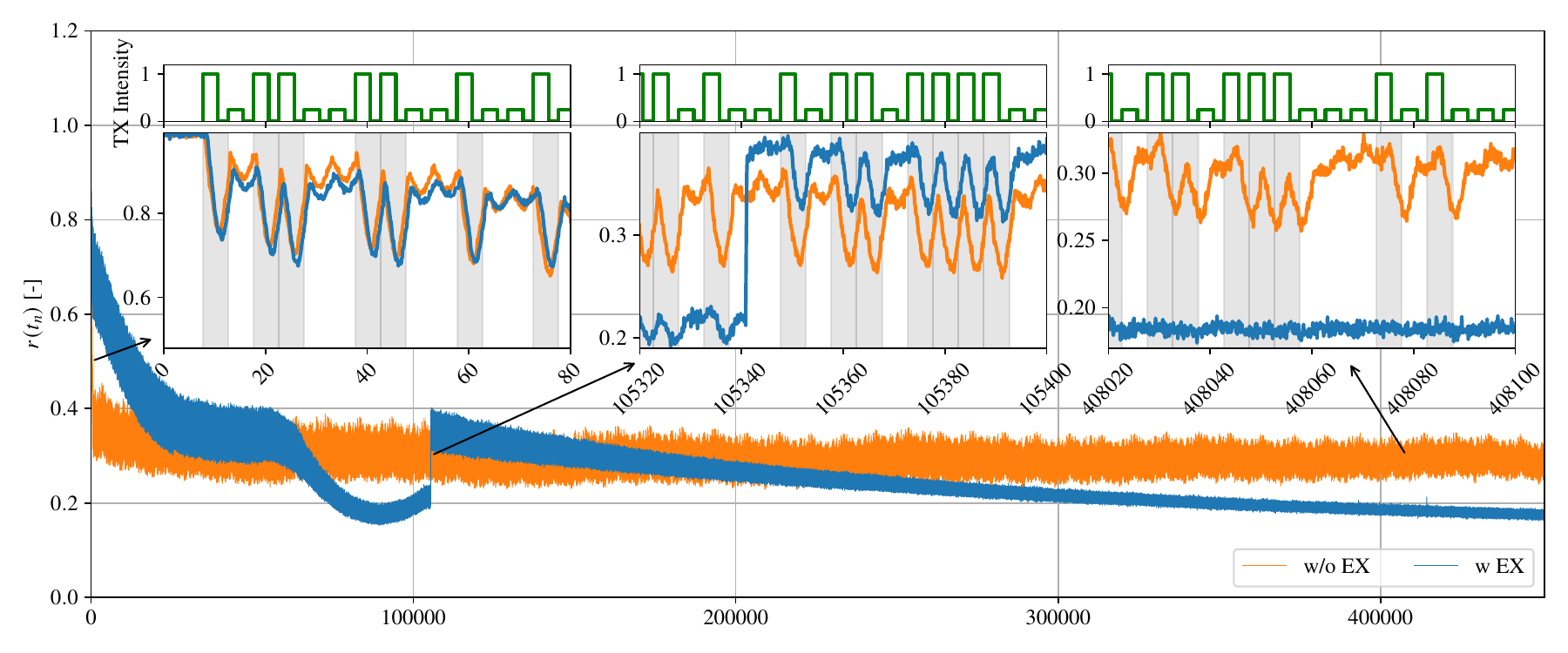}
        \label{fig:long_term_time}
    \end{subfigure}\\[1ex]

    \begin{subfigure}[b]{1\textwidth}
        \caption{}
        \includegraphics[width=\textwidth]{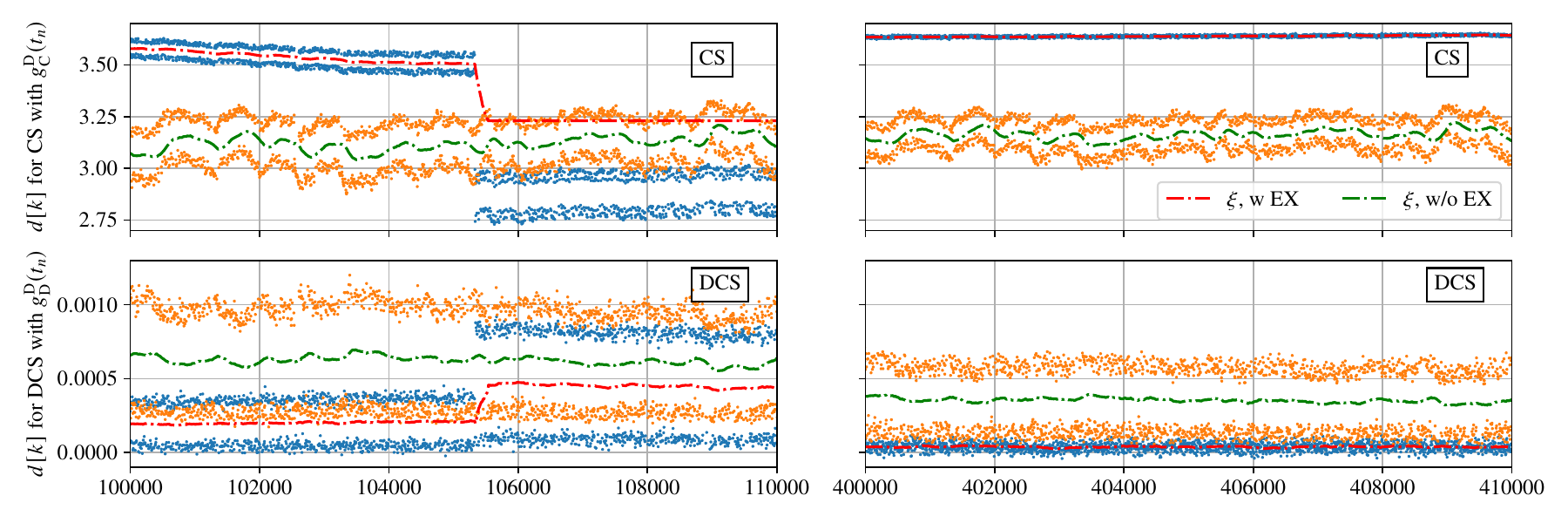}
        \label{fig:long_term_correlation}
    \end{subfigure}\\[1ex]

    \begin{subfigure}[b]{1\textwidth}
        \caption{}
        \includegraphics[width=\textwidth]{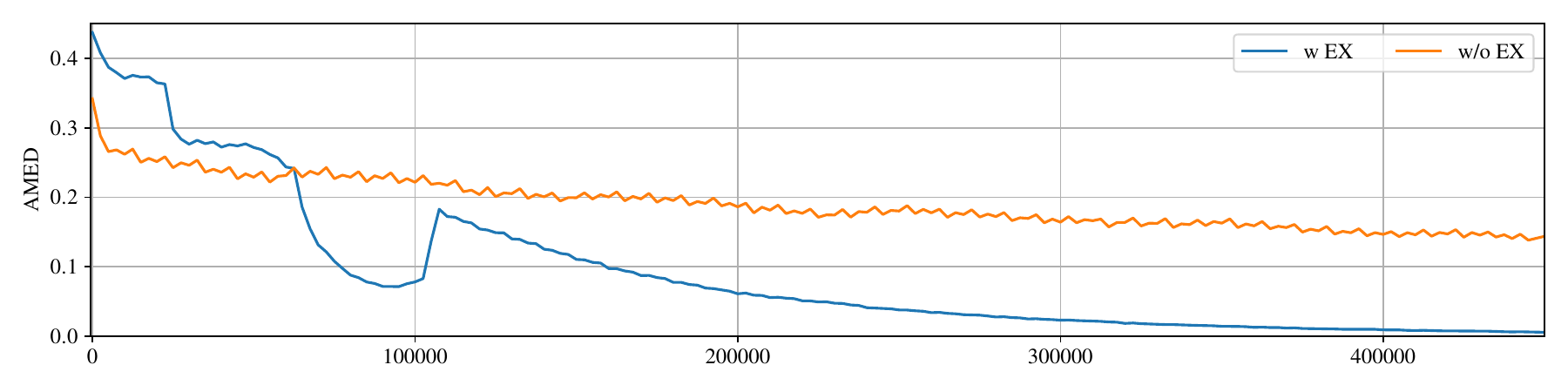}
        \label{fig:long_term_med}
    \end{subfigure}

    \caption{Evaluation of a long-term binary transmission with $\Ts = \SI{5}{\s}$, $\nBits = \num{90000}$, $\pFA = 10 \, \times \, 10^{-10}$, $\Delta t = \SI{0.1}{\s} $, $\filterLength = \lfloor \frac{\Ts}{\Delta t} \rfloor$, $\trainLength = 50$, $r = 0.04$, $W = 50$, $F = 1$, $\nSkip = 80$, and $P = 130$ with (blue curves) and without (orange curves) \ac{EX}. (a) Received signal over time. Shaded areas in the insets correspond to bit 1 transmissions. (b) Detection samples for different detection schemes using data-based receive filters and the corresponding adaptive thresholds, cf. \Sections{sec:init_thresholds}{sec:adaption_algorithm}, around $\SI{27}{\hour}$ after the start of data transmission (left panel) and at the end of transmission (right panel). (c) \ac{AMED} between bit 1 and bit 0 transmissions over time.}
    \label{fig:longterm}
\end{figure*}

\begin{figure*}[!tbp]\ContinuedFloat
    \centering

    \begin{subfigure}[b]{1\textwidth}
        \caption{}
        \includegraphics[width=\textwidth]{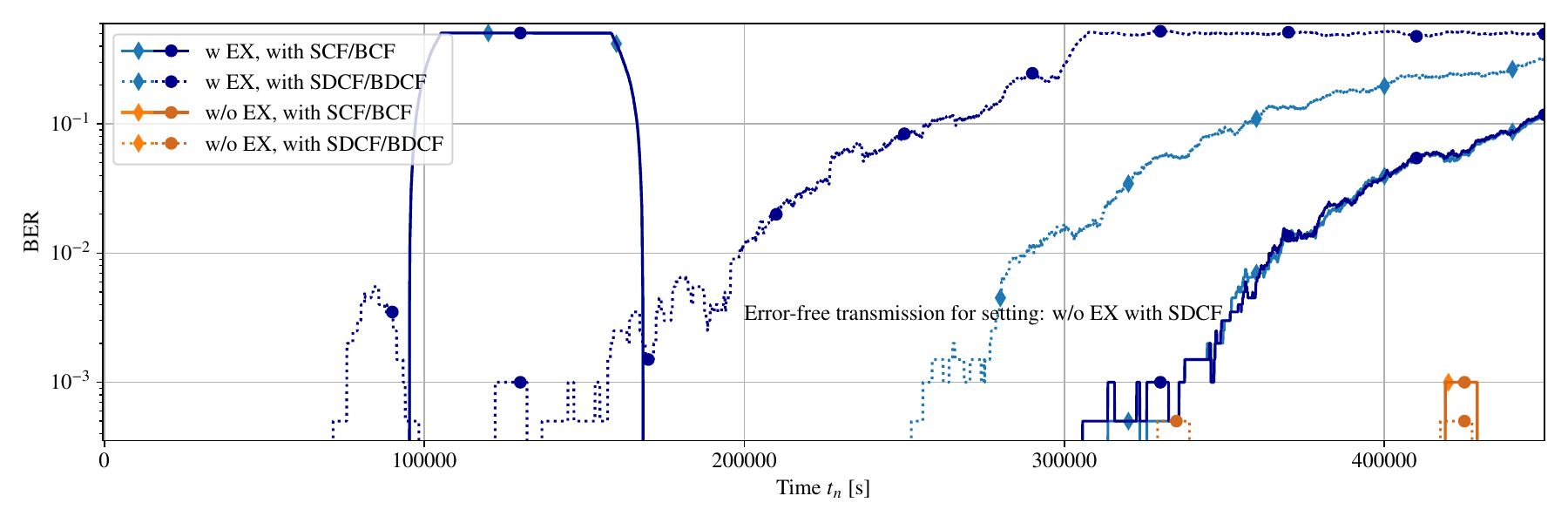}
        \label{fig:long_term_ber}
    \end{subfigure}

    \caption{(d) Moving average \ac{BER} (computed over the most recent $\num{2000}$ bit) for the different detection schemes.}
    \label{fig:longterm_part2}
\end{figure*}
In this section, we evaluate the performance of the testbed for long-term transmissions. The results are provided in \Figure{fig:longterm}. To this end, two experiments, one with \ac{EX} and one without \ac{EX}, were conducted over a period of 125 hours each, during which, in each case, $\nBits = \num{90000}$ were transmitted using binary modulation at a data rate of $\SI{12}{\bitsperminute}$. The $\num{90000}$ bit were generated using a random $\num{2000}$ bit sequence, which was repeatedly transmitted $45$ times. These experiments aim to demonstrate the key advantage of the proposed self-contained and closed-loop testbed, namely enabling the long-term study of tube-based MC systems. During each experiment, the testbed was not subjected to any alterations. Specifically, neither new \ac{GFPD} molecules were introduced nor were any substances extracted from the testbed. Instead, the same $9 \,\si{\milli\liter}$ of \ac{GFPD} solution was employed throughout each experiment.
\subsubsection{Received Fluorescence Signal}\label{sssec:time_signal}
\Figure{fig:long_term_time} shows the received signal $r(\tn)$ over time for both experiments with (blue curve) and without (orange curve) \ac{EX}. The inset plots show the signals at three particularly interesting time instants: the beginning (left), the time of an (undesired and unexpected) fluorescence jump in one of the received signals (center), and the end of the transmission (right), alongside with the corresponding \ac{TX} signal (green). Here, gray shaded areas in the inset plots indicate bit "1" transmissions. The inset plots of \Figure{fig:long_term_time} show that, as expected, $r(\tn)$ exhibits large drops in fluorescence intensity for bit 1 transmissions and small drops in fluorescence intensity for bit 0 transmissions. This indicates that reliable communication is possible. We further observe from \Figure{fig:long_term_time} that the general fluorescence level of the system decreases over time. This is due to offset \ac{ISI} and photobleaching effects, i.e., the prevalence of temporarily off-switched and permanently degraded \acp{GFPD}, respectively. We observe that, as expected, in the beginning ($t < \SI{60000}{\s}$), the average fluorescence level with \ac{EX} is higher than without \ac{EX}. However, later, the fluorescence for the case with \ac{EX} clearly drops below that without \ac{EX}. This indicates that the \ac{EX} successfully mitigates offset \ac{ISI}, which is the predominant immediate cause of a fluorescence decrease. At the same time, the \ac{EX} enhances photobleaching as its use increases the number of photons that hit the \ac{GFPD} molecules. The signal depicted in blue in the right inset plot reveals the increased photobleaching with \ac{EX}, which results in a performance degradation. In particular, in this case, the number of \ac{GFPD} molecules in the ON state that are available for off-switching at the \ac{TX} is very low. The modulation at the \ac{TX} thus exhibits fluorescence drops that cannot be discerned from measurement noise. This is not the case for the scenario without \ac{EX}.

One peculiar feature of the considered experiment is the rapid fluorescence decrease ($t \approx \SI{60000}{\s}$) and subsequent jump ($t \approx \SI{110000}{\s}$) for the scenario with \ac{EX}. We believe that this behavior is caused by an agglomeration of degraded \ac{GFPD} molecules in the flow cell, i.e., the \ac{RX}, as degraded proteins are known for their increased tendency to aggregate, cf. \Section{par:photobleaching}. Since degradation is caused by photobleaching, we expect this agglomeration will be more likely to occur in scenarios with \ac{EX} than without \ac{EX}. An agglomeration can significantly alter the measured fluorescence signal, as it acts as an undesirable physical light filter. This hypothesis matches the observed behavior, as a spontaneous detachment of the agglomeration would explain the re-establishment of the previous fluorescence level at $t \approx \SI{110000}{\s}$.
\subsubsection{Detection Samples}\label{sssec:detection_samples}
\noindent \Figure{fig:long_term_correlation} shows the detection samples, cf. \Equation{eq:detection_sample}, obtained using the \ac{CS} (top panels) and \ac{DCS} schemes (bottom panels) alongside the respective adaptive thresholds. We focus on two particularly interesting time intervals: The left panels focus on the time around the fluorescence jump, while the right panels show the samples at the end of the transmission. Orange and blue dots correspond to the detection samples, cf. \Equation{eq:detection_sample}, for the setting without and with \ac{EX}, respectively. First, we observe from the upper left segment of \Figure{fig:long_term_correlation} that as long as the range of detection samples $\detecSample[\symIdx]$ does not change rapidly, the threshold values $\detecThresh$ successfully track the temporal variations in $\detecSample[\symIdx]$, i.e., the adaption algorithm, cf. \Sections{sec:init_thresholds}{sec:adaption_algorithm}, operates successful. In particular, we observe that the threshold for the \ac{DCS} scheme (bottom) quickly adapts to the fluorescence jump, while errors are caused for the \ac{CS} scheme. The fluorescence jump, which predominantly results in an offset of the fluorescence, exerts a lower impact on the differential signal, which enables the adaptive algorithm to operative as expected. Moreover, \Figure{fig:long_term_correlation} confirms our prior observation that the experiment without \ac{EX} exhibits a more stable long-term behavior. For instance, we observe from the right-hand panels in \Figure{fig:long_term_correlation} that the bit 1 and bit 0 samples are still clearly separated by the threshold in the scenario without \ac{EX} even after $\SI{400000}{\s}$ ($\approx$ 111h) of continuous transmission, while for the scenario with \ac{EX}, the samples cannot be separated by either detection scheme. Interestingly, the advantage of the \ac{DCS} scheme vanishes at the end of the transmission for the experiment with \ac{EX}. The reason for this behavior is that the signal at this point is similar in magnitude to the noise. As the differentiation operation amplifies the noise, it starts to dominate the signal, which degrades the communication performance.
\subsubsection{AMED Analysis}\label{sssec:amed_long}
\noindent \Figure{fig:long_term_med} shows the \ac{AMED}, cf. \Equation{eq:amed}, for the scenarios with and without \ac{EX} over time. The metric is computed continuously over windows of $\nWindow = 500$ bit. As expected, the \ac{AMED} generally decreases over time, regardless of whether the \ac{EX} is used or not. Initially, when the offset \ac{ISI} is the dominant impairment, the use of the \ac{EX} leads to a higher \ac{AMED}. However, as photobleaching caused by the \ac{EX} becomes pronounced, the \ac{AMED} drops below that of the scenario without \ac{EX}. In addition, the effect of the fast fluorescence decrease and subsequent jump is clearly visible in \Figure{fig:long_term_med}. This is consistent with our previous observations for the fluorescence signal and the detection samples. Based on the \ac{AMED}, the benefits of the \ac{EX} for short binary transmissions become apparent, but they disappear for long sequences. In contrast, as we will discuss in \Section{ssec:eye_res}, for higher-order modulation schemes, the use of the \ac{EX} becomes essential as these schemes are considerably more sensitive to \ac{ISI}.
\subsubsection{BER Analysis}\label{sssec:ber_long}
Finally, \Figure{fig:long_term_ber} shows the empirical moving average \ac{BER}, i.e., the \ac{BER} calculated sequentially over the most recent $\num{2000}$ symbols using \Equation{eq:BER}, as a function of time for the different considered receive filters and the setting with and without \ac{EX}, respectively. Plotting the \ac{BER} this way enables us to analyze the causes of detection errors and whether the detectors allow for an adaptation to avoid subsequent errors or not. In \Figure{fig:long_term_ber}, the solid and dotted lines correspond to the use of the \ac{CS} and \ac{DCS} schemes, respectively, while the color indicates whether or not the \ac{EX} was active. The diamond-shaped and round markers specify whether data-based or blind filters were used. We observe that, while the \ac{EX} is in principle beneficial for short transmission times, for long transmission times, the \ac{BER} is larger for the case with \ac{EX}. This is, on the one hand, caused by the abnormal fast fluorescence decrease and fluorescence jump, and on the other hand, by the more severe photobleaching and the resulting low fluorescence level leading to a low signal-to-noise ratio. The former manifests itself in the sudden jump in the \ac{BER} to $0.5$ for the \ac{CS} scheme around the occurrence of the fluorescence jump. Detection errors end when the fluorescence is back roughly to the level at which it was before the jump, such that the threshold becomes valid again. The decreasing signal-to-noise ratio manifests itself in a gradual increase of the \ac{BER} over time. Here, the \ac{BER} rises first for the \ac{DCS} scheme, which can be attributed to the noise enhancement caused by differentiation. In contrast, without \ac{EX}, transmission remains error-free for more than $\SI{300000}{\s}$ for both considered detection schemes. In fact, error-free transmission of around $\nBits = \num{90000}$ bit\footnote{To be precise, with $\nSkip = 80$ and $\nPilots = 130$, $\nBits = 89790$ bit were detected error-free.} is possible with the \ac{DCS} scheme and the data-based receive filter $g^{\mathrm{D}}_{\mathrm{D}}(\tn)$.
{\sisetup{detect-all} 
\subsection{BER vs. Data Rate: 8-ary Modulation with an Achievable Rate of \textit{\SI{36}{\bit \per \minute}}}\label{sec:higher_order_modulation_rocks}}
\begin{figure*}[!tbp]
    \centering
    \includegraphics[width =1\textwidth]{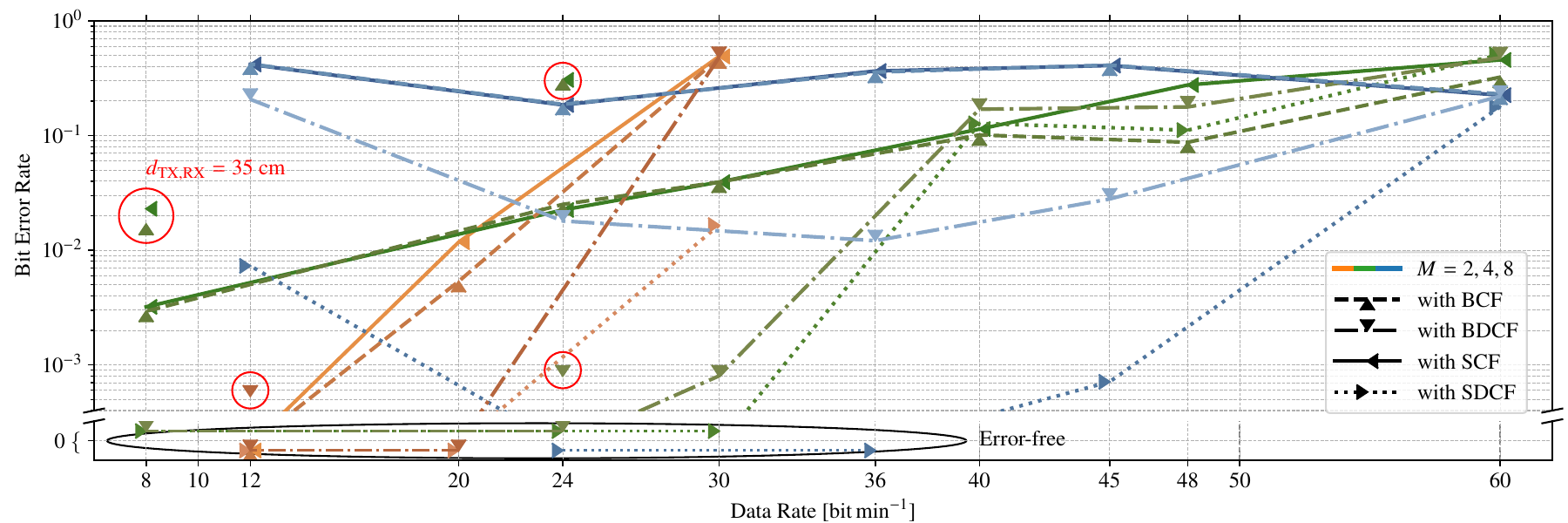}
    \caption{\acp{BER} for 18 individual experiments with a transmission length of $2000$ symbols each. Here, $\pFA = 10 \, \times \, 10^{-8}$, $\Delta t = \SI{0.1}{\s} $, $\filterLength = \lfloor \frac{\Ts}{\Delta t} \rfloor$, $\trainLength = 50$, $W = 50$, $F = 1$, $\nSkip = 80$, $P = 130$, and different search radius lengths $r \in [0.03, 0.12]$ are used. Thus, in each experiment, $1790 = 2000 -\nSkip - P$ symbols are used to determine the \ac{BER}, i.e., the \acp{BER} correspond to averages taken over $1790$, $3580$, or $5370$ bit, depending on the modulation order used. The red circles mark results obtained for $d_{\mathrm{TX},\mathrm{RX}} = 35 \,\si{\centi\meter}$.}
    \label{fig:ber}
\end{figure*}
We evaluate our testbed in terms of the achievable \ac{BER} for various different data rates. Specifically, we experimentally determined the \ac{BER} for 18 individual experiments, in each of which $\num{2000}$ symbols (corresponding to $\num{2000}$, $\num{4000}$, or $\num{6000}$ bit, depending on the modulation order) were transmitted. For each single experiment, a fresh \ac{GFPD} sample was used in order to prevent dependencies between the experiments. For all results shown in this section, the \ac{EX} was used. As we use $\nSkip = 80$ and $P = 130$, the \ac{BER} values were determined based on $1,790$ symbols.

\Figure{fig:ber} shows the achieved \acp{BER} for data rates between $\SI{8}{\bitsperminute}$ and $\SI{60}{\bitsperminute}$, cf. \Equation{eq:data_rate}. In particular, modulation orders $M = 2$ (yellow), $M = 4$ (green), and $M = 8$ (blue), and symbol durations between $\Ts = \SI{2}{\s}$ and $\Ts = \SI{15}{\s}$ are used. The individual experiments thus correspond to transmission durations between $\SI{67}{\minute}$ and $\SI{500}{\minute}$. Furthermore, we present results for a transmission channel length of $d_{\mathrm{TX},\mathrm{RX}} = 35 \,\si{\centi\meter}$ (marked by red circles) to provide insight into the effect of the channel length and to prove that reliable communication is still feasible for longer channels. Experiments for which no errors were observed are displayed at the bottom of \Figure{fig:ber} after the axis break.
\subsubsection{Effect of Modulation Order}\label{sssec:effect_modulation_order}
\Figure{fig:ber} shows that, for all considered modulation orders examined, error-free transmission was experimentally achieved within a certain data rate region (for $M=2$ from $\SI{12}{\bitsperminute}$ up to $\SI{20}{\bitsperminute}$, for $M=4$ from $\SI{8}{\bitsperminute}$ up to $\SI{30}{\bitsperminute}$, and for $M=8$ from $\SI{24}{\bitsperminute}$ up to $\SI{36}{\bitsperminute}$).
For $M=2$ and $M=4$, we observe that the \ac{BER}, as expected, increases with increasing data rate. For $M=8$, when comparing the \acp{BER} for $\SI{24}{\bitsperminute}$ and $\SI{12}{\bitsperminute}$, we see that the \ac{BER} is larger at $\SI{12}{\bitsperminute}$, which is not intuitive. However, when analyzing the received signal for this setting\footnote{Unfortunately, the mentioned received signal cannot be shown in this paper due to space constraints. However, of course, the data for this received signal is included in the dataset that we have made publicly available on Zenodo, cf. \Section{ssec:datasharing}.}, we observed a gradual decrease in fluorescence intensity, similar to that described in \Section{sssec:time_signal}. We attribute this to the severe photobleaching when using $\Ts = \SI{15}{\s}$ with irradiation duration $\Ti = \SI{10}{\s}$. Therefore, the resulting weak received signal does not support error-free transmission for 8-ary modulation for this data rate.

Note that, for a given modulation order $M$, we increase the data rate by reducing the symbol duration $\Ts$. However, $\Ts$ cannot be reduced arbitrarily, as this leads to an increase in channel \ac{ISI}, which in turn results in an increased \ac{BER}. Specifically, \Figure{fig:ber} reveals that only $M=8$ can achieve error-free transmission at rates greater than $\SI{30}{\bitsperminute}$. Hence, for error-free transmission at high data rates, an increase in modulation order is necessary. For such a setting, where $M=8$ and $\Ts = \SI{5}{\s}$ are used, details are provided next.

\begin{figure*}[!tbp]
\centering
  \includegraphics[width = \textwidth]{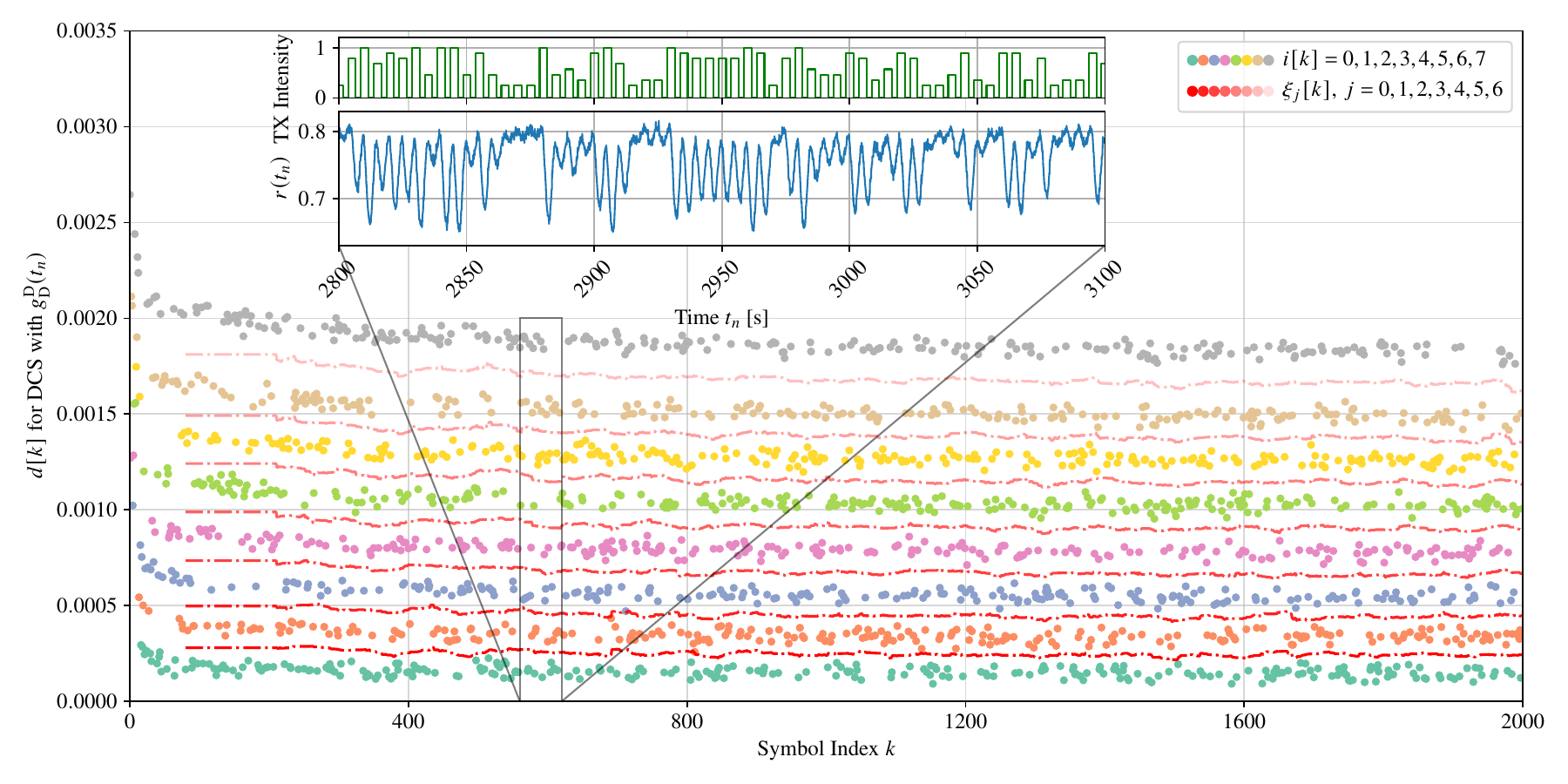}
  \caption{Detection samples as defined in \Equation{eq:detection_sample} and adaptive thresholds over time. Here, $\modOrder = 8$, $\Ts = \SI{5}{\s}$, $\Delta t = \SI{0.1}{\s}$, $\filterLength = \lfloor \frac{\Ts}{\Delta t} \rfloor$, $\trainLength = 50$, $W = 50$, $F = 1$, $\nSkip = 80$, $P = 130$, and $r = 0.08$ are used. The detection samples are color-coded based on the corresponding transmit symbol $\symVar[k] \in \{0, \ldots, M - 1\}$. Inset plot: Corresponding received signal and \ac{TX} intensity.}
  \label{fig:M8_detection}
\end{figure*}
\Figure{fig:M8_detection} shows the detection samples $\detecSample[\symIdx]$ for the \ac{DCS} scheme using \ac{SDCF} as receive filter. In \Figure{fig:M8_detection}, the different colors indicate the transmitted symbols $\symVar[k] \in \{0,\ldots,\modOrder - 1\}$, while the dash-dotted lines in the red monochrome color palette correspond to the adaptive thresholds $\modOrder - 1$ used for symbol detection. Additionally, the inset in \Figure{fig:M8_detection} shows the \ac{TX} intensity and the received signal for the symbols transmitted in the interval $t \in [2800, 3100] \, \si{\s}$.

 From the inset in \Figure{fig:M8_detection}, we observe that the \ac{TX} intensity is reflected in visible changes of the received signal, which shows distinct peaks. We further observe from \Figure{fig:M8_detection} that, even for 8-ary modulation, the samples for the \ac{DCS} scheme are well distinguishable \ac{wrt} the adaptive threshold values. This is attributed to the use of the guard interval and the \ac{EX} as well as the employed detection scheme. In fact, although noise and \ac{ISI} are present in the received signal $r(t)$, they are effectively suppressed through a combination of differentiation and matched filtering. Specifically, the filtering described in \Equation{eq:detection_sample} suppresses the noise. Additionally, we employ \ac{DCS}, i.e., $\Tilde{\vec{\recSig}} (\tsEst[\symIdx])$ in \Equation{eq:detection_sample} corresponds to the sampled differential received signal $\Tilde{\recSig}(\tn) = \diffRecSig(\tn) = \recSig(t_{\n+1}) - \recSig(\tn)$, cf. \Section{sec:symbol_sync}. This differential approach effectively mitigates offset \ac{ISI}. Moreover, \Figure{fig:M8_detection} shows that the experiment is in a transient phase at the beginning, which leads to a high variance of the sample values. The transient phase is caused by offset \ac{ISI}, which develops slowly over successive transmissions and remains relatively constant after about 50 to 100 symbols. Therefore, the omission of the first $\nSkip = 80$ symbols is crucial to ensure that the thresholds $\detecThresh_\detecThreshIdx$ can be adequately initialized afterwards.
\subsubsection{Effect of Receive Filter and Detection Method}\label{sssec:effect_filter_and_detection}
In \Figure{fig:ber}, we can also compare the \acp{BER} for the four filters proposed for detection: \ac{SCF} (solid, leftwards pointing marker), \ac{BCF} (dashed, upwards pointing marker), \ac{SDCF} (dotted, rightwards pointing marker), and \ac{BDCF} (dash-dotted, downwards pointing marker). Detection based on the differential signal, i.e., \ac{SDCF} and \ac{BDCF}, in many cases leads to a significantly lower \ac{BER} in direct comparison to \ac{SCF} and \ac{BCF}. Furthermore, \Figure{fig:ber} shows that, as expected, the data-based filters (\ac{SCF} and \ac{SDCF}) are superior to the blind filters (\ac{BCF} and \ac{BDCF}), as evidenced by the lower \ac{BER}.

We conclude that the \ac{DCS} scheme is well suited for detection of higher-order modulation when used in combination with \ac{EX}. However, the \ac{DCS} scheme can be disadvantageous in long experiments, where the signal eventually becomes weak due to photobleaching, because of the noise enhancement introduced by the differentiation operation. In this case, the \ac{CS} scheme is preferable, cf. \Section{sec:error_free_80kbit}.
\subsubsection{Effect of Transmission Distance}\label{sssec:effect_distance}
Finally, \Figure{fig:ber} shows the effects of the elongation of the channel from $d_{\mathrm{TX},\mathrm{RX}} = 6 \,\si{\centi\meter}$ to $d_{\mathrm{TX},\mathrm{RX}} = 35 \,\si{\centi\meter}$. Three distinct settings have been analyzed: $M=4$ with data rate $\SI{8}{\bitsperminute}$, $M=2$ with data rate $\SI{12}{\bitsperminute}$, and $M=4$ with data rate $\SI{24}{\bitsperminute}$. For these settings, error-free transmission is still possible even for $d_{\mathrm{TX},\mathrm{RX}} = 35 \,\si{\centi\meter}$ when using the \ac{SDCF} as receive filter\footnote{Note that only the markers for $\textnormal{BER}>0$ are shown with the red circle. Error-free results for a distance of $35 \,\si{\centi\meter}$ are not indicated by a marker to avoid overloading the figure around $\textnormal{BER}=0$.}. Moreover, we observe from \Figure{fig:ber} that for the detection schemes, where detection errors occur, the corresponding errors are larger for distance $d_{\mathrm{TX},\mathrm{RX}} = 35 \,\si{\centi\meter}$ compared to $d_{\mathrm{TX},\mathrm{RX}} = 6 \,\si{\centi\meter}$, which was expected. For example, for $M=4$ and the \ac{CS} scheme, the \ac{BER} increases by one order of magnitude.
\subsection{EX-Based ISI Mitigation in Short-Term Transmissions}\label{ssec:eye_res}
\begin{figure*}[!tbp]
    \centering
    \begin{tabular}{@{} c @{\hspace{1em}} c @{\hspace{1em}} c @{}}
        & \small{$\disTX = \SI{6}{\cm}$} & \small{$\disTX = \SI{35}{\cm}$} \\
        \raisebox{1.75\height}{\rotatebox{90}{\small{w \ac{EX}}}} &
        \includegraphics[width=0.45\textwidth]{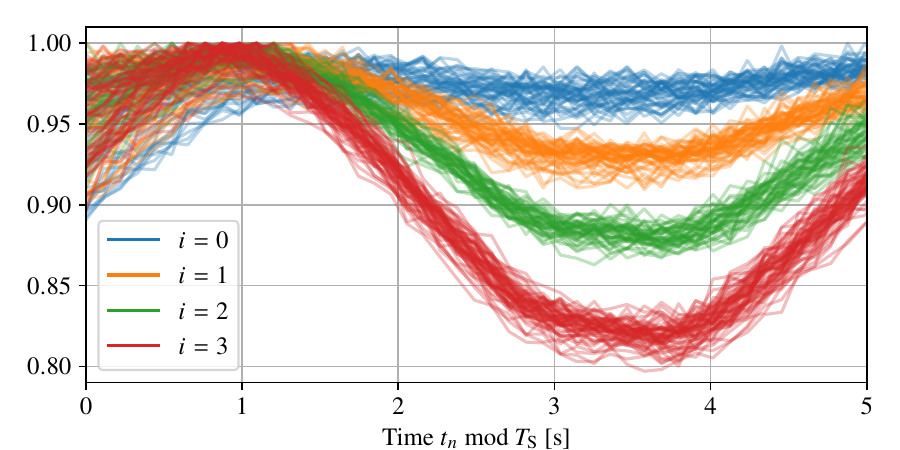} &
        \includegraphics[width=0.45\textwidth]{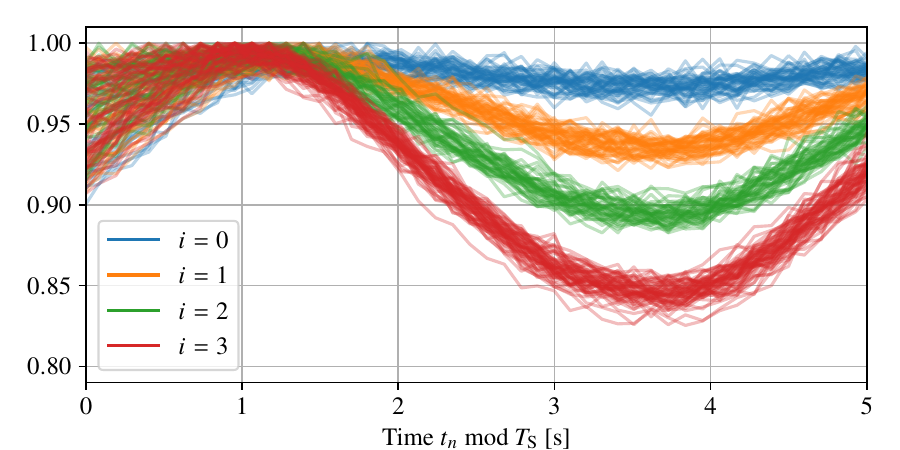} \\
        \raisebox{1.25\height}{\rotatebox{90}{\small{w/o \ac{EX}}}} &
        \includegraphics[width=0.45\textwidth]{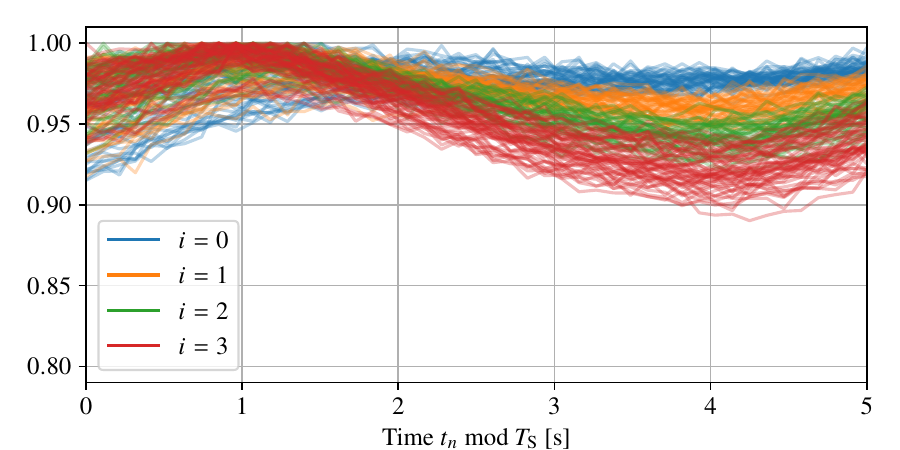} &
        \includegraphics[width=0.45\textwidth]{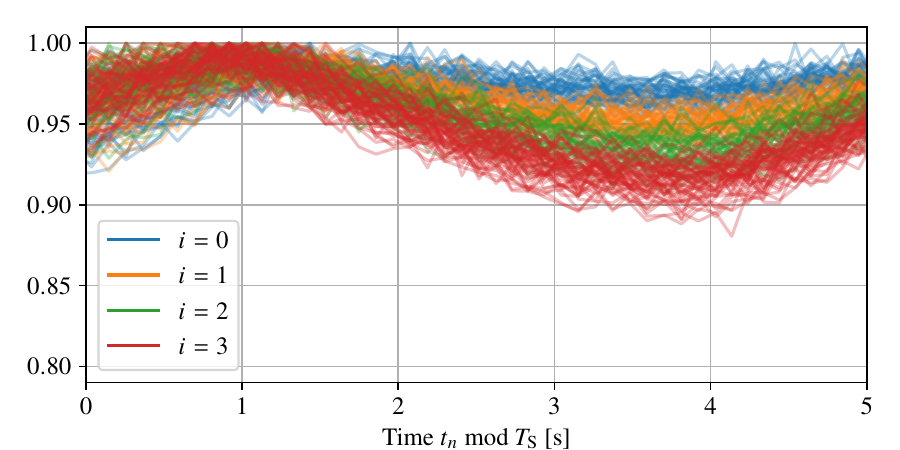}
    \end{tabular}
    \caption{Eye diagrams. Here, $\modOrder=4$, $\Ts = \SI{5}{\s}$, $\Ti = \SI{3}{\second}$, and $\Delta t = \SI{0.1}{\s}$ are used.}
    \label{fig:eye_diagram}
\end{figure*}
In this section, we show that the \ac{EX} is capable of \ac{ISI} mitigation in short-term transmission. \Figure{fig:eye_diagram} shows eye diagrams for $\modOrder = 4$ and different experimental settings, including varying channel lengths and \ac{EX} settings. The curves correspond to symbols $\symVar[k]$ for $k \in  \{100, 101, \ldots, 299\}$ out of a $\nSymb = \num{2000}$ symbol long sequence being transmitted. Synchronization was achieved using the \ac{CS} scheme with \ac{SCF} as receive filter for all scenarios. We horizontally shifted the received signal $r(\tn)$ for each transmit symbol to have its maximum at 1 for better visibility. In \Figure{fig:eye_diagram}, different transmit symbols $\symVar$ are denoted by different colors.

When the \ac{EX} is used (top panels in \Figure{fig:eye_diagram}), we observe three eyes, indicating that the different transmit symbols can be distinguished easily. This is especially evident for the short channel length of $\disTX = \SI{6}{\centi\meter}$ (top left panel). Elongating $\disTX$ results in a slightly smaller eye opening. Thus, the larger $\disTX$ has only a small effect, which shows that, as analyzed theoretically in \Section{subsec:propagation}, diffusion affecting the TX-RX link can be neglected for the $\disTX$ considered in this work.

If the \ac{EX} is turned off, we see that the eyes are closed (bottom panels in \Figure{fig:eye_diagram}), i.e., the received signals for the different transmit symbols $\symVar$ are largely overlapping.
In particular, we observe that the received signals for the same symbol $\symVar$ vary more without \ac{EX} compared to when the \ac{EX} is used, which we attribute to larger inter-loop and offset \ac{ISI}. We observe from \Figure{fig:eye_diagram} that for the considered settings, reliable communication using 4-ary modulation without \ac{EX} can be difficult and may require sophisticated equalization techniques at the \ac{RX}.

In summary, employing an \ac{EX} has some benefits (\ac{ISI}-mitigation, cf. \Section{ssec:eye_res}), but also undesired side-effects (photobleaching of \ac{GFPD}, cf. \Section{sec:error_free_80kbit}).
Based on our results, we conclude that the use of an \ac{EX} is advisable for short-term transmission, high modulation orders, and short symbol durations $\Ts$.
\section{Testbed Comparison and Data Sharing}\label{sec:comparison_Testbeds}
In this section, we compare our testbed to other fluid-based testbeds, and provide information regarding the sharing of our data and code.
\subsection{Comparison with Other Testbeds}\label{ssec:comparison}
\begin{table*}[!tbp]
    \caption{Overview of Selected Fluid-Based \ac{MC} Testbeds}
    \centering
    \resizebox{1\textwidth}{!}{
        \begin{tabular}{cccccc}
            \hline\hline
             Signaling Molecule (Biocompatibility) & Data rate (bit/s) & \ac{BER} (\# transmitted bits) & Detection Method & Reference\\
             \hline\hline
             Green Fluorescent Protein Dreiklang (GFPD) (\cmark) & 0.6 & 0 (5370) & Differential Signal + Adaptive Threshold & This paper\\
             \hline
             Superparamagnetic Iron Oxide & 10 & 0.0843 (600) & Convolutional Neural Network &  \cite{bartunik2023development}\\
             Nanoparticles (SPIONs) (\cmark) &  &  &  & \\ 
             Sodium chloride (NaCl) (\cmark)& 5 & 0.002 (100 x 100) & $\mu$ Link Decoder with Channel Estimation & \cite{wang2020understanding}\\
             & &  & and Sequence Detection &  \\
             Colored Ink (\xmark) & 1.5 & 1/312 (312) & Threshold & \cite{wietfeld2024evaluation}\\ 
             Acid/Base (\xmark) & 2.63 & 0 (1080) & Recurrent Neural Network & \cite{farsad2017novel}\\
             Sodium Hydroxide (NaOH) (\xmark) & 1/375 & 0.04 (100) & Threshold &  \cite{walter2023real}\\
             Hydrogen Chloride (HCl) (\xmark) & 1/27 & 0.022 (1000) & Adaptive Threshold & \cite{khaloopour2019experimental}\\
             Transfer DNA (tDNA) (\cmark) & 1/120 & 0 (20) & Differential Signal + Threshold & \cite{kuscu2021fabrication}\\
             Glucose ($\textrm{C}_6 \textrm{H}_{12} \textrm{0}_6$) (\cmark) & 2 & 0.05 (654) & Machine Learning & \cite{koo2020deep}\\
             \hline\hline
        \end{tabular}
    }
    \label{tab:experiments}
\end{table*}
Our comparison is focused on the experimentally determined \acp{BER}, the detection schemes used, and the resource efficiency of data transmission. \Table{tab:experiments} provides a selected overview of fluid-based \ac{MC} testbeds, listing for each testbed the signaling molecule used, the lowest reported \ac{BER}, the data rate achieving this \ac{BER}, the length of the bit sequence transmitted, and the employed detection method.

Providing a fair and still meaningful comparison between existing \ac{MC} testbeds and the proposed testbed is challenging due to some fundamental differences.
In particular, none of the existing works has considered a closed-loop and self-contained topology. As a result, only our testbed is confronted with the challenge of resolving inter-loop \ac{ISI}, offset \ac{ISI}, and permanent \ac{ISI}. In addition, all other testbeds require and use a much larger number of signaling molecules due to their open topology. The resulting low resource efficiency \ac{wrt} molecule usage, i.e., molecule efficiency, which in \cite{bartunik2021increasing} has been defined as the number of bit transmitted per volume of dissolved signaling molecule solution, is a significant challenge in existing testbeds. The issue of low molecule efficiency has been recognized and discussed \cite{bartunik2023development, koo2020deep, lin2024testbed}: In \cite{bartunik2023development}, an efficiency of $160 \,\si{\bit \per \milli\liter}$ was achieved, which is still low compared to our testbed with $90,000/9 \, \si{\bit \per \milli\liter} = 10,000 \,\si{\bit \per \milli\liter}$; in \cite{koo2020deep}, the authors reported that some measurements had to be discarded because the \ac{TX} ran out of molecules earlier than expected; and in \cite{lin2024testbed}, high costs were associated with the use of the testbed, resulting in a maximum evaluated bit sequence length of $8$ bit\footnote{The authors in \cite{lin2024testbed} also reported that for such a short sequence a meaningful estimation of the \ac{BER} was not possible. We believe that this is an example where the discussed alternative performance metrics, such as \ac{AMED} and eye diagrams, are particularly useful.}.

Despite the different topologies and the challenges our testbed has to overcome, we achieve a data rate that is only an order of magnitude lower than the largest data rates reported in the \ac{MC} literature, which are from \cite{bartunik2023development}\footnote{For the sake of completeness, we note that, to the best of our knowledge, the highest \textit{achievable} data rate was reported in \cite{huang2024non}.}. At the same time, our testbed provides very reliable communication, i.e., very low \acp{BER}, while employing a low-complexity detection scheme that involves only differentiation of the received signal and an adaptive threshold detector. Quantifying the computational complexity of the detection schemes listed here is beyond the scope of this paper. However, methods based on machine learning, such as Convolutional Neural Networks \cite{bartunik2023development} and Recurrent Neural Networks \cite{farsad2017novel}, as well as methods that explicitly take into account the memory effects caused by \ac{ISI} \cite{wang2020understanding}, even if efficiently implemented using the Viterbi algorithm, are believed to have comparatively higher computational complexities.
\subsection{Data Handling and Code Sharing}\label{ssec:datasharing}
To promote transparency and enable experimental evaluation of new communication algorithms developed by the \ac{MC} community, we publish our experimental data and the Python code for synchronization and detection in Zenodo and a Git repository under the CC BY and the MIT licenses, respectively. When used, the data and/or code can and should be cited using the corresponding Zenodo \ac{DOI} \cite{scherer2025Zenodo}. The link to the Git repository can be found on Zenodo.

\section{Conclusion}\label{sec:conclusion}
In this paper, we presented the first self-contained, closed-loop experimental \ac{MC} system using media modulation. As the testbed leverages the reusable and biocompatible signaling molecule \ac{GFPD}, long-term experiments without the need for repeated injection or removal of molecules were possible.

We developed a communication scheme which features higher order modulation, a noise-based wake-up method, blind and data-based synchronization, and adaptive threshold detection. Utilizing key performance metrics, such as \ac{AMED} and \ac{BER}, the quality and reliability of the testbed were demonstrated. Notably, we achieved error-free transmission of $\num{5370}$ bit at $36 \, \si{\bitsperminute}$. Moreover, we conducted the longest \ac{MC} experiment to date, both \ac{wrt} the number of bits transmitted as well as the duration of the transmission. In particular, $\num{90000} \, \si{bit}$ were transmitted error-free over a period exceeding $5$ days ($125 \, \si{\hour}$) at a data rate of $12 \, \si{\bitsperminute}$, thereby setting a novel benchmark for long-term MC experiments. In total, only $9 \, \si{\milli \liter}$ of \ac{GFPD} solution were required for this long experiment, which serves to illustrate the efficacy of the media modulation approach.
In order to encourage further research, we have made the experimental data, which corresponds to the received samples collected for the \ac{MC}-based transmission of over 250 kbit of information, and the corresponding evaluation code available via open access on Zenodo and Github, respectively. The availability of the raw experimental data enables researchers to explore advanced synchronization and detection schemes (e.g., based on machine learning) to improve communication performance, which are interesting topics for future research.
Moreover, the testbed can easily be expanded due to its modular and flexible design. For example, the incorporation of a branched tube network, which could emulate real biological environments, is of high practical interest. In addition, to better align with biological applications, exploring the miniaturization of the used communication system components, such as the light-based \ac{TX}, \ac{EX}, and \ac{RX}, represents a promising direction for future experimental research.
In the long term, we expect that extensions of the proposed closed-loop testbed will enable testing of \ac{IoBNT} interfaces, evaluation of drug delivery mechanisms, and investigation of \ac{MC} between implanted devices. These applications are designed for use in the human cardiovascular system, which features a closed molecular transport loop - a key feature that our testbed reproduces.

\bibliographystyle{IEEEtran}
\bibliography{literature}
\section*{Acknowledgment}
{\small
We thank Prof. Stefan Jakobs (Max Planck Institute for Biophysical Chemistry, Göttingen, Germany) for providing a plasmid encoding Dreiklang. This work was funded by the Deutsche Forschungsgemeinschaft (DFG, German Research Foundation) under Project-ID 290825040 and GRK 2950 – Project-ID 509922606, and by the Horizon Europe Marie Skodowska-Curie Actions (MSCA)-UNITE under project 101129618.
}
\end{document}